\documentclass[iop]{emulateapj}

\usepackage{braket}

\shorttitle{Chondrule Formation and Planetary Accretion}
\shortauthors{Hasegawa et al}

\begin{document}

\title{Chondrule Formation via Impact Jetting Triggered by Planetary Accretion}

\author{Yasuhiro Hasegawa\altaffilmark{1,3}, Shigeru Wakita\altaffilmark{2}, Yuji Matsumoto\altaffilmark{2}, Shoichi Oshino\altaffilmark{2}}
\affil{$^1$Division of Theoretical Astronomy, National Astronomical Observatory of Japan, Osawa, Mitaka, Tokyo 181-8588, Japan}
\affil{$^2$Center for Computational Astrophysics, National Astronomical Observatory of Japan, Osawa, Mitaka, Tokyo 181-8588, Japan}
\email{yasuhiro.hasegawa@nao.ac.jp}

\altaffiltext{3}{EACOA fellow}

\begin{abstract}
Chondrules are one of the most primitive elements that can serve as a fundamental clue as to the origin of our Solar system.
We investigate a formation scenario of chondrules that involves planetesimal collisions and the resultant impact jetting.
Planetesimal collisions are the main agent to regulate planetary accretion 
that corresponds to  the formation of terrestrial planets and cores of gas giants.
The key component of this scenario is that ejected materials can melt 
when the impact velocity between colliding planetesimals exceeds about 2.5 km s$^{-1}$.
The previous simulations show that the process is efficient enough to reproduce the primordial abundance of chondrules.
We examine this scenario carefully by performing semi-analytical calculations 
that are developed based on the results of direct $N$-body simulations.
As found by the previous work,
we confirm that planetesimal collisions that occur during planetary accretion can play an important role in forming chondrules.
This arises because protoplanet-planetesimal collisions can achieve the impact velocity of about 2.5 km s$^{-1}$ or higher,
as protoplanets approach the isolation mass ($M_{p,iso}$).
Assuming that the ejected mass is a fraction ($F_{ch}$) of colliding planetesimals' mass, 
we show that the resultant abundance of chondrules is formulated well by $F_{ch}M_{p,iso}$,
as long as the formation of protoplanets is completed within a given disk lifetime.
We perform a parameter study and examine how the abundance of chondrules and their formation timing change.
We find that the impact jetting scenario generally works reasonably well for a certain range of parameters,
while more dedicated work would be needed to include other physical processes that are neglected in this work
and to examine their effects on chondrule formation.
\end{abstract}


\section{Introduction} \label{intro}

Chondrules are one of the most primitive materials that 
probably contain profound information of how our Solar system formed \citep[e.g.,][for a most recent review]{dac14}.
As they are named, 
chondrules are the most abundant ingredient in chondrites (up to 80 \% in volume).
Chondrules have a number of the interesting properties; 
they are millimeter-sized particles, formed as molten droplets of silicate that has the melting temperature of around $1800$ K.
These properties provide invaluable constraints on their formation mechanisms.
In fact, intensive experiments reveal that 
the cooling rate of $10-1000$ K per hour is required to form chondrules \citep[e.g.,][]{jl93,hcl05}.
This implies that chondrules would probably have formed in the gas-rich phase of the solar nebula, rather than in empty space.
Furthermore, the age of chondrules indicates that 
chondrules started forming when calcium-aluminum-rich inclusions (CAIs) formed,
and since then chondrules kept forming for about three million years \citep[e.g.,][]{cbk12}.
As CAI formation began about 4567 million years ago \citep[e.g.,][]{akl10,cbk12},
it is generally anticipated that understanding of how chondrules formed may 
shed light on origins of our Solar system.

A number of formation mechanisms of chondrules have been proposed so far \citep[e.g.,][for a review]{dmc12}.
These include X-wind models \citep[e.g.,][]{ssl96,ssg01}, solar nebula lighting \citep[e.g.,][]{msl93,dc00}, 
and nebular shocks excited by planetesimals in eccentric orbits \citep[e.g.,][]{w96,ins01,md10}.
While different models focus on different physical mechanisms,
the main motivation of these models is to re-construct the thermal history of chondrules;
the progenitor of chondrules is melted first and then the resultant liquid of silicate cools down rapidly to form spherical droplets.
It is currently recognized that none of them can fully reproduce a wealth of chondrules' properties.

\citet{jmm15} have recently performed a series of numerical simulations 
and shown that collisions among planetesimals and the resultant impact jetting may be a plausible candidate to form chondurles.
It is interesting that the idea of planetesimal collisions was initially proposed in 1960's \citep[e.g.,][]{uc53}.
At that time, however, it was thought that collisions did not play an important role in forming chondrules,
because the collisional outcome is fragmentation of colliding bodies, rather melting them.
One of new findings achieved by \citet{jmm15} is that, 
when planetesimals collide with each other, some materials can be ejected from their coillisonal surface.
This process is referred to as impact jetting and was indeed suggested by \citet[][]{ss12}.
\citet{jmm15} have for the first time performed numerical simulations of impact jetting, 
and shown that ejected materials can reach the melting temperature of chondrules
when the impact velocity of planetesimals exceeds about 2.5 km s$^{-1}$.
They have thus found that impact jetting triggered by planetesimal collisions can serve as an essential process 
to reproduce a number of the key properties of chondrules including their abundance.
Since terrestrial planets and cores of gas giants can be formed via planetesimal collisions \citep[e.g.,][for recent reviews]{rkm14,hbp14},
which is often referred to as planetary accretion,
their work finally suggests that the formation of chondrules is a natural consequence of planet formation.

\citet{jmm15} have adopted a rather sophisticated technique to model planetary accretion,
wherein collisions among planetesimals and between planetesimals and growing protoplanets 
are simulated by means of a Monte Carlo approach \citep[also see][]{ml14}.
The technique is obviously needed to accurately trace a collisional history of planetesimals,
which enables a better estimate of the chondrule mass formed by impact jetting.
As its very nature, however, stochasticity modeled by the Monte Carlo method may end up with masking of
how individual (or combined) physical processes come into play in generating the resultant outcome,
and hence has some potential to jeopardize its importance.

Here, we re-visit the scenario proposed by \citet{jmm15}, and examine it in detail.
To simplify our analysis, 
we perform semi-analytical calculations that are developed based on more detailed, direct $N-$body calculations.
We consider two kinds of collisions that can occur in planetary accretion;
protoplanet-planetesimal collisions and planetesimal-planetesimal ones.
By computing the impact velocity of colliding planetesimals that is determined 
both by the random velocity of small planetesimals and by the surface escape velocity,
we will show below that protoplanet-planetesimal collisions play a major role in forming chondrules.
This is an outcome that the surface escape velocity becomes more important as a protoplanet grows.
We confirm the trend by estimating the total mass of chondrules that can be formed by impact jetting via these two kinds of collisions.
Furthremore, we find that, when planetary accretion is complete within a given disk lifetime,
the total mass of chondrules is well characterized by the isolation mass,
and that its value can exceed the mass of the current astroid belt at $a\simeq 2-3$ AU.
We also examine the formation timescale of chondrules by impact jetting,
and show that, for a reasonable range of disk mass and lifetime,
the resultant value can match the timescale that is implied from the currently available samples.
As found by \citet{jmm15}, therefore,
impact jetting is a promising process to understand an origin of chondrules,
while more extensive work would be desired to fully calibrate the resultant abundance of chondrules and 
the corresponding timing of chondrule formation.

The plan of this paper is as follows.
In Section \ref{model}, we briefly discuss the current theory of planetary accretion, and summarize its formulation.
Based on the formulation, we discuss how chondrule-forming impacts can be realized in planetary accretion (see Section \ref{res}).
We consider both protoplanet-planetesimal collisions and planetesimal-planetesimal ones.
We also explore some parameter space to examine the dependence of our results on some parameters.
In Section \ref{disc}, we discuss implications of impact jetting for chondrule formation,
and consider other physical processes that are not included in this work.
Conclusive remark of this work is given in Section \ref{conc}.

\section{Theory of Planetary Accretion} \label{model}

We begin with an introduction to the current view of planetary accretion \citep[e.g.,][]{ki00,rkm14}.
Here, a brief summary is developed by simply describing the key quantities that involve chondrule-forming impacts.
Table \ref{table1} tabulates important valuables.
{Readers who are familiar with planetary accretion can go directly to Section \ref{res},
where its application to chondrule formation is discussed.

\begin{table*}
\begin{minipage}{17cm}
\begin{center}
\caption{Summary of key quantities}
\label{table1}
\begin{tabular}{clc}
\hline
Symbol           &   Meanings                                           & Value  \\ \hline 
$\rho_{gas}$   &   Gas density at the disk midplane &            \\
$\Sigma_{d}$  &   Dust surface density                    &            \\
$f_d$               &   Increment factor of $\Sigma_d$                          & 1 or 3  \\ 
$\tau_d$          &   Disk lifetime  (as an upper limit)               &     $10^7$ yr \\    \hline
$M_p$            &   Mass of runaway bodies/protoplanets &   \\
$R_p$            &    Radius of runaway bodies/protoplanets &   \\
$\rho_p$        &    Mean density of runaway bodies/protoplanets & 5 g cm$^{-3}$  \\
$v_M$             &  Random velocity of runaway bodies/protoplanets &   \\
$a$                 &  Orbital position of runaway bodies/protoplanets &   \\ \hline
$m_{pl}$         &   Mass of field planetesimals           & $10^{23}$ g, $10^{24}$ g     \\
$\rho_{pl}$        &  Mean density of field planetesimals & 2 g cm$^{-3}$  \\
$r_{pl}$           &  Radius of field planetesimals           & $\simeq$ 230 km, 500 km \\
$v_m$             &  Random velocity of field planetesimals &   \\  \hline
$v_{rel}$        &   Relative velocity between a planetesimal and a planetesimal/runaway body/protoplanet & $\simeq v_{m}$ \\
$v_{esc}$      &  Surface escape velocity of planetesimals      &  \\    
$v_{imp}$      &  Impact velocity of planetesimals                    & $\sqrt{v_m^2 + v_{esc}^2}$ \\
$v_{Kep}$        &  Keplarian velocity &   \\
$C$                 &  Accretion acceleration factor                 &  2 \\
$M_{p,iso}$     & Isolation mass of protoplanets                  & \\
$\langle e_{eq}^2 \rangle_{i}^{1/2} $ & Root mean square equilibrium eccentricity & \\  
                       &      of field planetesimals in the $i$ growth mode       &   \\ \hline
$v_c$             &   Value of $v_m$ required for chondrule formation  & $\simeq$ 2.5 km s$^{-1}$ \\              
$e_c$             &    Eccentricity required for chondrule formation ($\simeq v_c / v_{Kep}$)     &       \\   \hline
$\tau_{int}$    & Time when $v_{imp} = v_c$                                                                      &      \\
$\tau_{end}$  & Time when growing protoplanets obtain $M_{p,iso}$                               &      \\
$F_{pl} (v_{imp} > v_c)$ & Fractional number of planetesimals that satisfy $v_{imp} > v_c$ &      \\
$F_{ch}$              &  Mass fraction of planetesimals that can eventually generate chondrules via impact jetting  & $10^{-2}$  \\
$\Delta m_{ch,M-m}$  &     Cumulative mass of chondrules formed via protoplanet-planetesimal collisions    &     \\
$\Delta m_{ch,m-m}$  &     Cumulative mass of chondrules formed via planetesimal-planetesimal collisions    &     \\ 
$\Delta m_{ch}$             &  Cumulative mass of chondrules formed via impact jetting over the disk lifetime &  $\Delta m_{ch,M-m} + \Delta m_{ch,m-m} $  \\
\hline
\end{tabular}
\end{center}
\end{minipage}
\end{table*}

\subsection{Definition of planetesimals' velocities}

The velocity ($v_{rel}$) of a planetesimal relative to other planetesimals in a swarm of planetestimals
can be given as \citep{ls93}
\begin{equation}
 v_{rel} \equiv \left( \frac{5}{4} e^2 + i^2 \right)^{1/2} v_{kep},
\end{equation}
where $e$ and $i$ is the eccentricity and the inclination of the planetesimal, respectively.
On the other hand, the velocity ($v_{m}$) of a planetesimal relative to the mean circular orbit in the midplane
with the same semimajor axis as that of the planetesimal, 
can be written as
\begin{equation}
 v_{m} \equiv \left( e^2 + i^2 \right)^{1/2} v_{kep}.
\end{equation}
Assuming the equipartition of the random energy, that is, $\langle e^2 \rangle^{1/2} \simeq 2 \langle i^2 \rangle^{1/2}$,
these two equations give
\begin{equation}
\label{eq:v_m}
v_{rel} \simeq v_{m} \simeq \langle e^2 \rangle^{1/2} v_{Kep}.
\end{equation}
In the following section, we assume that the random velocity of planetesimals 
is described approximately as $v_{m}$.
Note that we have confirmed that this choice does not affect our results.

\subsection{Disk Model}

Before going into the theory of planetary accretion,
we introduce a disk model that serves as a basis for the following discussion.

We adopt a conventional model that has been widely used in the currently existing $N$-body simulations.
The model is essentially comparable to the famous minimum-mass solar nebular (MMSN) model \citep[e.g.,][]{h81},
and can be written as (see Table \ref{table1}),
\begin{equation}
\rho_{gas} = 2 \times 10^{-9} f_d \left( \frac{a}{1 \mbox{ AU}} \right)^{-11/4} \mbox{ g cm}^{-3},
\end{equation}
and
\begin{equation}
\Sigma_d = 10 f_d  \left( \frac{a}{1 \mbox{ AU}} \right)^{-3/2} \mbox{ g cm}^{-2},
\end{equation}
where $f_d$ is an increment factor to examine the effect of disk mass on planetary accretion.
For the standard case, we adopt $f_d=1$, while $f_d=3$ is also considered as a massive disk case, 
following \cite{jmm15}.
We consider the case that the stellar mass is $M_* = 1 M_{\odot}$.

\subsection{Runaway \& Oligarchic Growth}

It is well known that 
planetary accretion that corresponds to the growth of protoplanets via planetesimal collisions 
can divide into two stages: the so-called runaway and oligarchic growth.

Runaway growth occurs as the initial step of forming protoplanets \citep[e.g.,][]{ws89,ki96}.
In this phase, "runaway" bodies emerge out of a planetesimal disk, 
which is the outcome of gravitational focusing ($v_{rel} < v_{esc}$) and dynamical friction ($v_M < v_m$) \citep[e.g.,][]{oin93,ki96}.
Based on equation (\ref{eq:v_m}),
the growth rate of runaway bodies is regulated predominantly by the eccentricity of smaller planetesimals ($\langle e^2 \rangle^{1/2}$).
Assuming the balance between viscous stirring of planetesimals that pumps up their eccentricity and 
gas drag that damps their eccentricity, 
the equilibrium eccentricity in this phase ($\langle e_{eq}^2 \rangle_{run}^{1/2}$) is given as 
\citep[e.g.,][]{ki00}
\begin{eqnarray}
\label{eq:e_run}
\langle e_{eq}^2 \rangle_{run}^{1/2}  & \simeq &  4.3 \times 10^{-3} \left( \frac{m_{pl}}{10^{23} \mbox{ g}} \right)^{4/15}
                                               \left( \frac{\Sigma_d}{10 \mbox{ g cm}^{-2}} \right)^{1/5} \\ \nonumber
                           & \times &    \left( \frac{\rho_{gas}}{2 \times10^{-9} \mbox{ g cm}^{-3}} \right)^{-1/5}
                                               \left( \frac{a}{1 \mbox{ AU}} \right)^{1/5},
\end{eqnarray}
where laminar gas disks are assumed \citep{ahn76}.
Note that,  as described in Section \ref{intro},
gas disks would be needed to reproduce the cooling rate of chondrules \citep[e.g.,][]{jl93,hcl05}.
We also note that $\langle e_{eq}^2 \rangle_{run}^{1/2}$ roughly goes as $m_{pl}^{1/3}$.
It is important that this growth mode leads to a situation that 
a small number of larger planetesimals grow more rapidly than a large number of smaller planetesimals.
In the end, the mass of smaller planetesimals remains very small.
For simplicity, it is assumed that the mass of field planetesimals does not change with time in this paper.

Oligarchic growth is the second step of planetary accretion \citep[e.g.,][]{im93,ki98,ki02,tdl03}.
In this stage, runaway bodies become massive enough to disturb the motion of the surrounding small planetesimals.
We refer to such massive bodies as protoplanets.
As a result, the eccentricity of small planetesimals is pumped up by protoplants (rather than by the planetesimals themselves), 
and their system becomes dynamically heated up.
This ends up with gravitational focusing that is less effective than in the runaway phase.
Nonetheless, dynamical friction is still active, 
and hence the growth rate of protoplanets is regulated mainly by the eccentricity of small planetesimals (see equation (\ref{eq:v_m})).
Under the assumption that perturbations acting on planetesimals by a protoplanet become equal to
the damping effect by gas drag, 
the equilibrium eccentricity ($\langle e_{eq}^2 \rangle_{oli}^{1/2}$) is written as \citep[e.g.,][]{ki02}
\begin{eqnarray}
\label{eq:e_oli}
\langle e_{eq}^2 \rangle_{oli}^{1/2} & \simeq &  2.6 \times 10^{-2} \left( \frac{m_{pl}}{10^{23} \mbox{ g}} \right)^{1/15}
                                               \left( \frac{\rho_{pl}}{2 \mbox{ g cm}^{-3}} \right)^{2/15}  \\ \nonumber
                            & \times &  \left( \frac{\rho_{gas}}{2 \times10^{-9} \mbox{ g cm}^{-3}} \right)^{-1/5}
                                               \left( \frac{a}{1 \mbox{ AU}} \right)^{-1/5}   \\ \nonumber
                            & \times &   \left( \frac{M_{p}}{0.1 M_{\oplus}} \right)^{1/3},                  
\end{eqnarray}
where laminar gas disks are assumed again, and the normalized feeding zone of the protoplanet $\tilde{b} =10$ is adopted.
As pointed out in runaway growth, 
the mass of small planetesimals does not change significantly even in this phase,
and gas disks would be needed for chondrule formation.
It is also important to emphasize that, due to the presence of gas drag,
the random velocity of planetesimals is much slower than the escape velocity of protoplanets (i.e., $v_m < v_{esc}$).

\subsection{Mass Evolution of Protoplanets}

We are now in a position to compute the mass growth of protoplanets via planetary accretion.
Since the two-body approximation can be applied to both runaway and oligarchic growth \citep[e.g.,][]{oin93},
the change of protoplanets' mass with time can be written as \citep[e.g.,][]{im93}
\begin{equation}
 \label{eq:dmdt}
 \frac{dM_{p}}{dt} \simeq C \pi \Sigma_d \frac{2 G M_{p}R_p}{\langle e_{eq}^2 \rangle a v_{Kep}},
\end{equation}
where either $\langle e_{eq}^2 \rangle_{run}$ or $\langle e_{eq}^2 \rangle_{oli}$ will be substituted into $\langle e_{eq}^2 \rangle$, 
depending on a growth mode.
Note that $dM_p/dt$ decreases with increasing $\langle e_{eq}^2 \rangle$.
This is the main reason why the growth rate in the oligarchic phase is much slower than that in the runaway phase.

\begin{figure*}
\begin{minipage}{17cm}
\begin{center}
\includegraphics[width=8cm]{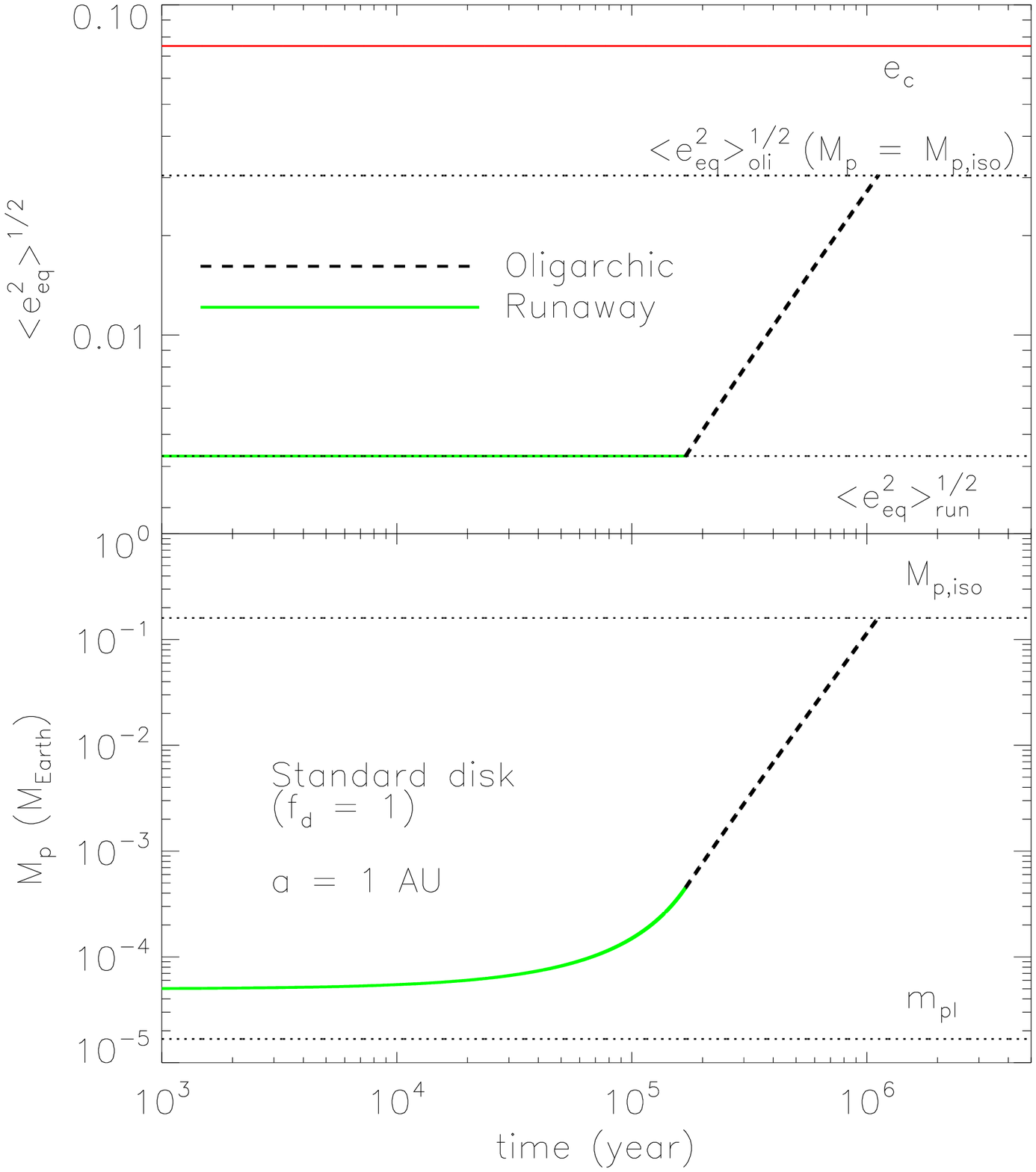}
\includegraphics[width=8cm]{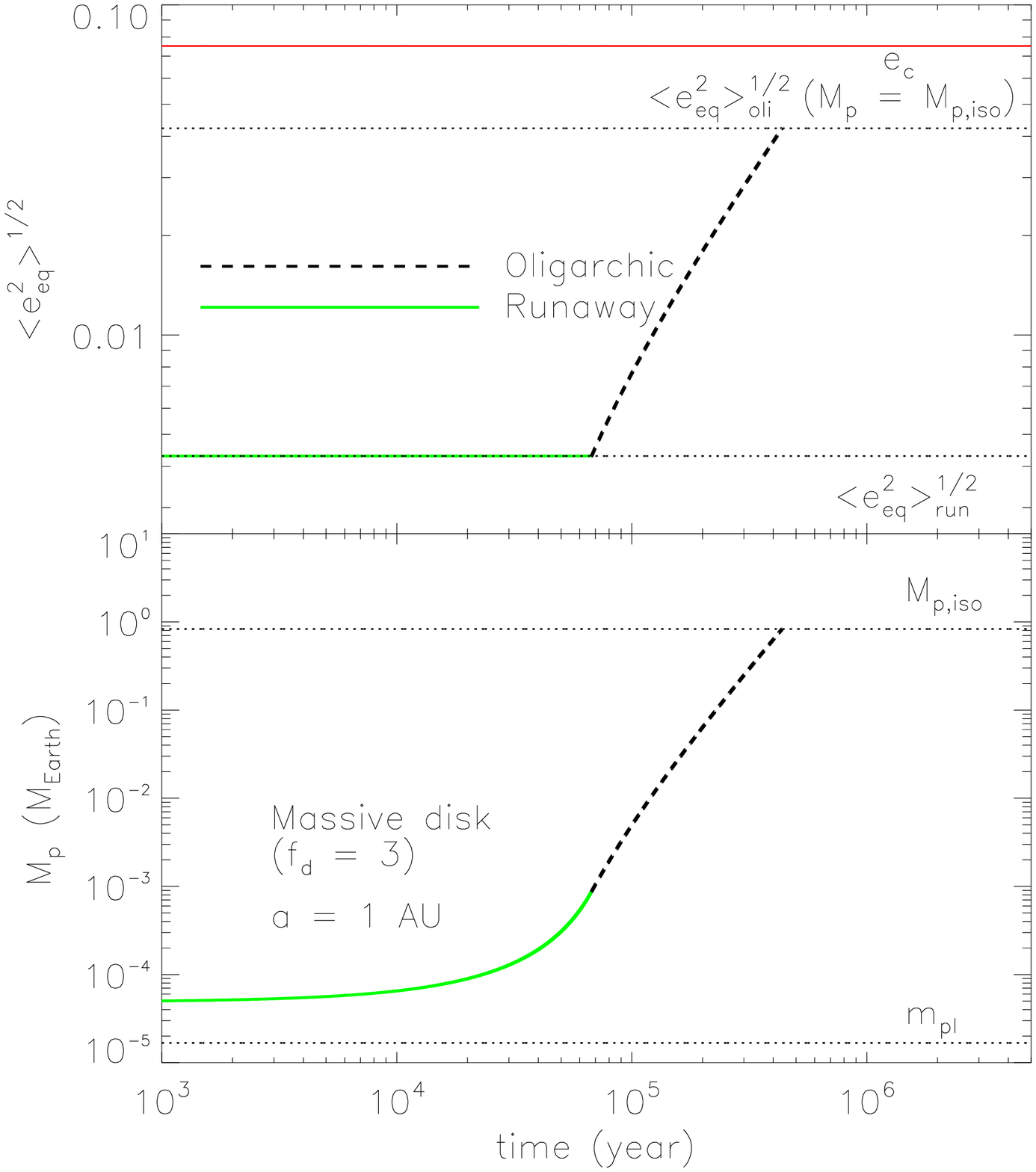}
\caption{Time evolution of $M_p$ and $\langle e_{eq}^2 \rangle^{1/2}$ at $a=1$ AU with $m_{pl} = 10^{23}$ g. 
The top panel shows $\langle e_{eq}^2 \rangle^{1/2}$ while the bottom panel is for $M_p$.
The results for the standard disk ($f_d=1$) are shown on the left panel,
and those for the massive disk ($f_d=3$) are on the right panel.
The increase of disk mass accelerates planetary growth and 
heats up the planetesimal disk (higher $\langle e_{eq}^2 \rangle^{1/2}$) when $M_p = M_{p,iso}$.
As a reference, the value of $e_c$ that is required for chondrule formation via impact jetting 
is plotted on the top panels.}
\label{fig1}
\end{center}
\end{minipage}
\end{figure*}

Figure \ref{fig1} shows the resultant behavior of $M_{p}$ and $\langle e_{eq}^2 \rangle^{1/2}$ that evolve with time, 
on the bottom and the top, respectively at the planetary position of $a = 1 $ AU.
The left panel shows the results of the standard case ($f_d=1$) while the right panel is for the massive case ($f_d=3$).
We adopt $m_{pl} = 10^{23}$ g.
As described above, runaway growth initially takes place (see the solid line) and 
$dM_p/dt$ is regulated by $\langle e_{eq}^2 \rangle_{run}^{1/2}$ (see equations (\ref{eq:e_run}) and (\ref{eq:dmdt})).
Note that $\langle e_{eq}^2 \rangle_{run}^{1/2}$ is constant with time 
since $\langle e_{eq}^2 \rangle_{run}^{1/2} \propto m_{pl}^{4/15}$ (see the top panels).
As $M_p$ increases with time (see the bottom panels), gravitational focusing becomes less effective.
When $\langle e_{eq}^2 \rangle_{oli}^{1/2}$ becomes roughly equal to $\langle e_{eq}^2 \rangle_{run}^{1/2}$,
perturbations by protoplanets become dominant (see the dashed line).
Equivalently, $M_p \simeq  50-100 \times m_{pl}$ \citep{im93}.
Then the growth mode switches from runaway growth to oligarchic one.
In the oligarchic phase, eccentricity evolution goes as $M_p^{1/3}$ (see equation (\ref{eq:e_oli})), 
and $dM_p/dt$ is computed, using $\langle e_{eq}^2 \rangle_{oli}^{1/2}$ (rather than $\langle e_{eq}^2 \rangle_{run}^{1/2}$).
Oligarchic growth continues
until protoplanets obtain the so-called isolation mass ($M_{p,iso}$) with which 
growing protoplanets consume all the planetesimals in their feeding zone.
In an actual formula, $M_{p,iso}$ is described as \citep[e.g.,][]{ki00}
\begin{eqnarray}
 \label{eq:mp_iso}
 M_{p,iso}  & \simeq &  0.16 M_{\oplus} 
                                     \left( \frac{\Sigma_d}{10 \mbox{ g cm}^{-2}} \right)^{3/2}  
                                     \left( \frac{a}{1 \mbox{ AU}} \right)^{3},
\end{eqnarray}
where we adopt $\tilde{b} =10$ again.
Our computation is terminated when $M_p$ reaches $M_{p,iso}$ (see the upper horizontal line on the bottom panels).
Our results show that planetary accretion at $a = 1$ AU is done well before a upper limit of disk lifetimes 
\citep[$\tau_d \simeq 10^7$ yr,][]{wc11}.
We have also performed $N$-body simulations using GRAPE computers, and
confirmed that the overall behavior of $M_{p}$ and $\langle e_{eq}^2 \rangle^{1/2}$ is 
generally consistent with the results of the simulations (Oshino et al. in prep).

There is essentially no qualitative difference between the standard ($f_d=1$) and the massive ($f_d=3$) cases.
Only noticeable features are that massive disks accelerate planetary growth (see equation (\ref{eq:dmdt})), and 
that the equilibrium eccentricity at $M_p = M_{p,iso}$ becomes a higher value 
(see equations (\ref{eq:e_oli}) and (\ref{eq:mp_iso})).  

\section{Chondrules formed by impact jetting} \label{res}

Chondrule formation via impact jetting requires a high value of the impact velocity ($v_{imp}$), 
which is about $v_c \simeq$2.5 km s$^{-1}$ \citep{jmm15}.
This is because ejected materials should melt to satisfy the formation condition of chondrules 
that is derived from the currently available samples.
In this section, we investigate under what condition(s) such collisions can be realized.
Since planetary accretion involves two kinds of collisions, 
namely, planetesimal-planetesimal collisions and protoplanet-planetesimal ones,
we compute their contribution to chondrule formation individually.
This enables an identification of when chondrule-forming impacts occur for each type of collisions,
and an estimate of which type of collisions plays a major role in forming chondrules.
We also compute the total mass of chondrules formed via planetary accretion 
and examine the dependence of the total mass on the value of $a$ and $m_{pl}$.
This is necessary to discuss how effective impact jetting is as a chondrule forming process.

\subsection{Definition of the impact velocity}

We first define the impact velocity ($v_{imp}$) of a planetesimal that undergoes impact jetting.
Assuming that a planetesimal with the mass of $m_{pl}$ and the radius of $r_{pl}$ is colliding with 
another body with the mass of $M_{i}$ and the radius of $R_{i}$, 
$v_{imp}$ can be given as (also see table \ref{table1})
\begin{equation}
\label{eq:v_imp}
v_{imp} \equiv \sqrt{ v_m^2 + v_{esc}^2,} 
\end{equation}
where the surface escape velocity ($v_{esc}$) is written as
\begin{equation}
v_{esc} = \sqrt{\frac{2G ( m_{pl} + M_i )}{r_{pl} + R_{i}}}.
\end{equation}
Note that $M_i =M_{p}$ and $R_i = R_p$ when protoplanet-planetesimal collisions are considered,
while $M_i = m_{pl}$ and $R_i = r_{pl}$ for planetesimal-planetesimal collisions.
Thus, we can effectively consider two kinds of collisions independently.

\subsection{Protoplanet-planetesimal collisions} \label{res_iso}

We investigate protoplanet-planetesimal collisions and their outcome on chondrule formation.
To proceed, we consider planetary accretion at $a= 1$ AU as done in Section \ref{model}.

\begin{figure*}
\begin{minipage}{17cm}
\begin{center}
\includegraphics[width=8cm]{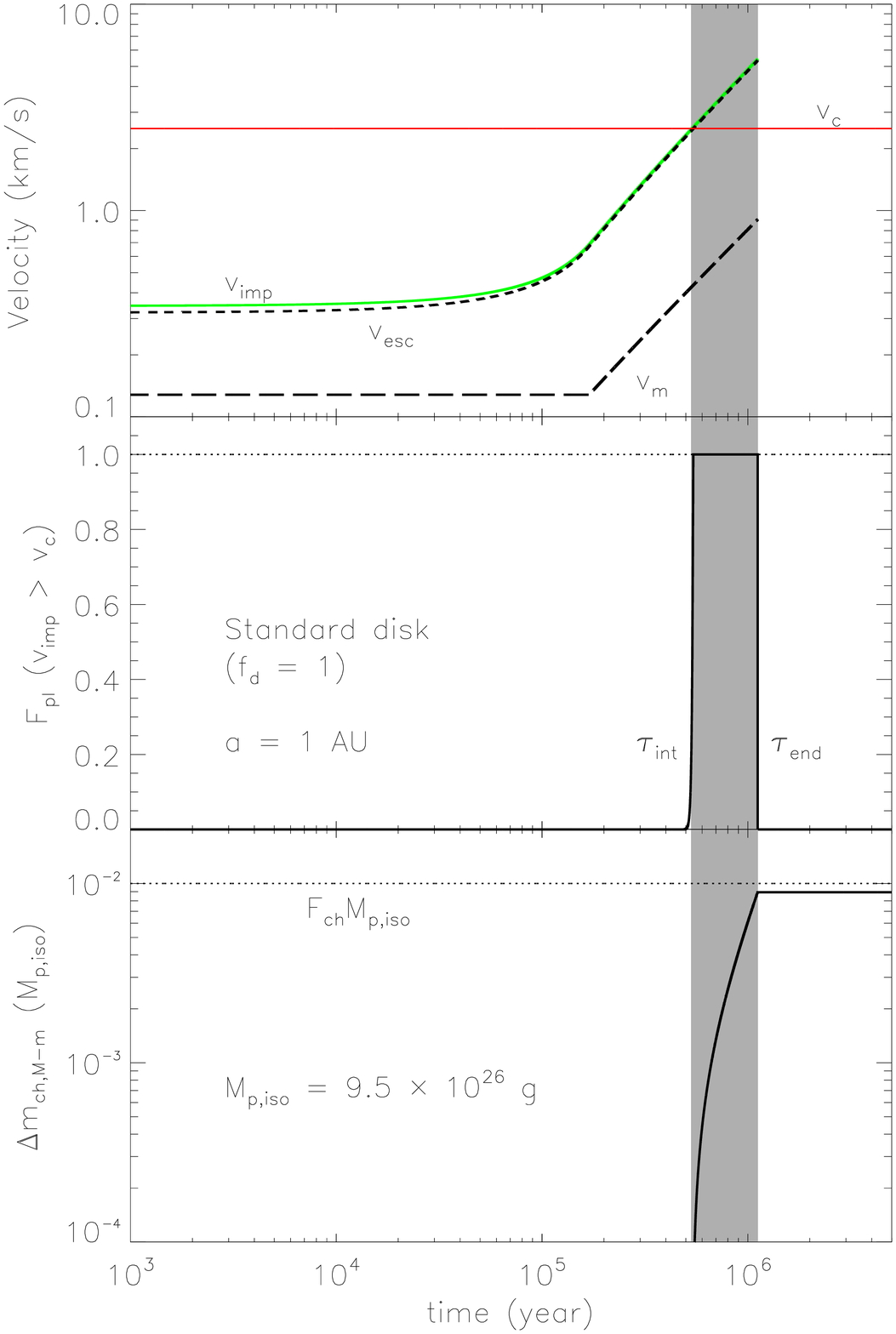}
\includegraphics[width=8cm]{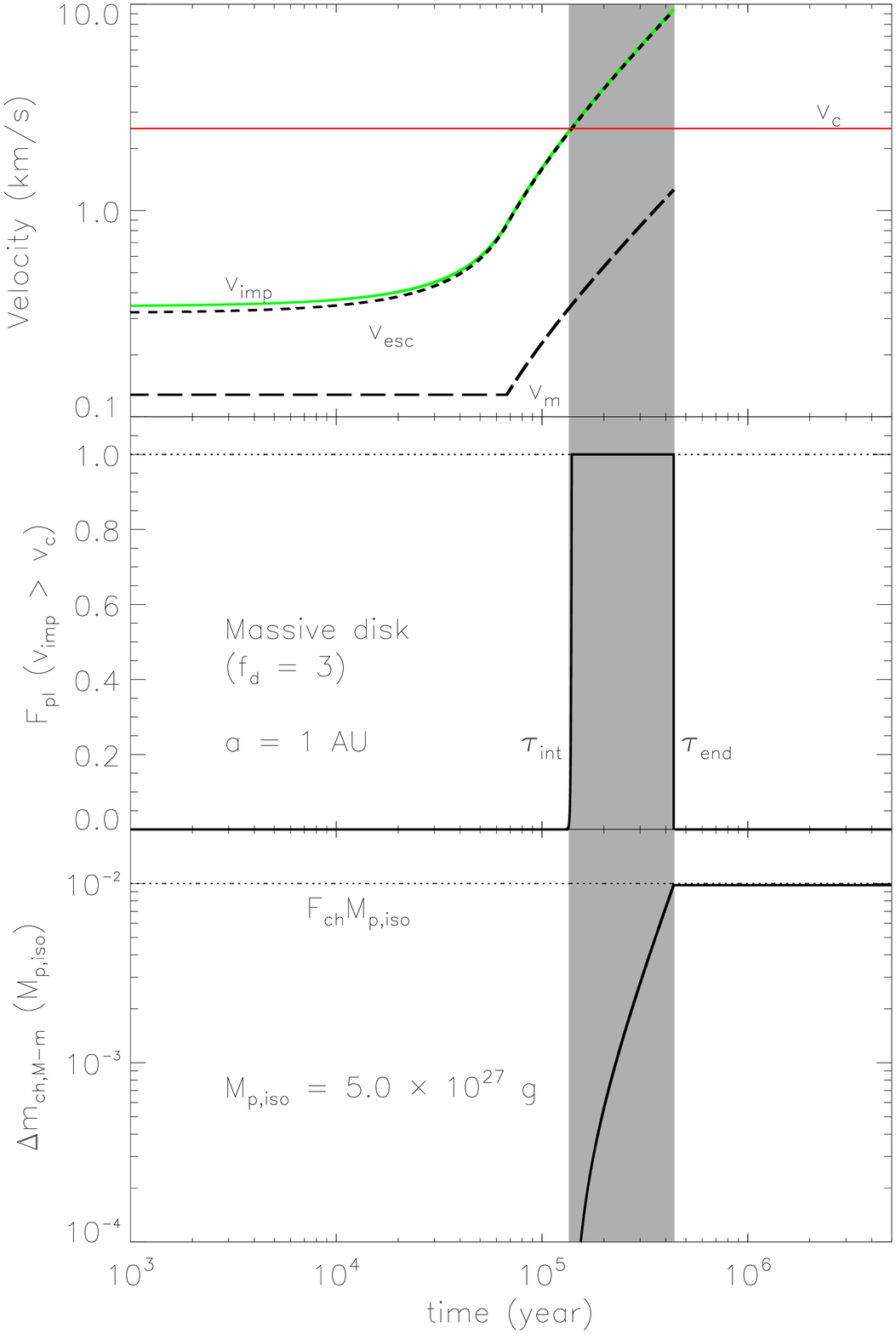}
\caption{The resultant impact velocity ($v_{imp}$), $F_{pl}(v_{imp} > v_c)$, and the cumulative chondrule mass ($\Delta m_{ch,M-m}$) 
as a function of time on the top, the middle, and the bottom, respectively, at the position of $a=1$ AU. 
Protoplanet-planetesimal collisions are considered.
The standard case ($f_d=1$) is shown on the left panel and the massive case ($f_d=3$) is on the right panel.
The value of $v_{imp}$ (the green, solid line) is determined largely by $v_{esc}$ (the dotted line), 
and can exceed $v_c$ (the red, solid line) at the later stage of planetary accretion (the top).
Both $v_{esc}$ and $v_{m}$ (the dashed line) behave similarly, because they are well coupled with each other via the growth of a protoplanet.
The value of $F_{pl}(v_{imp} > v_c)$ represents when chondrule-forming impacts can be realized (see the middle).
The majority of such impacts occur when $v_{imp}$ can exceed $v_c$ (i.e., $\tau_{int} < \tau < \tau_{end}$, also see the hatched region).
The resultant $\Delta m_{ch,M-m}$ is shown on the bottom panel.
As expected, chondrule formation takes place mainly at the hatched region.
It is important that the final value of $\Delta m_{ch,M-m}$ is approximately formulated by $F_{ch} M_{p,iso}$ (see the dotted line).}
\label{fig2}
\end{center}
\end{minipage}
\end{figure*}

Figure \ref{fig2} (top) shows the resultant value of $v_{imp}$ that evolves with time (see equation (\ref{eq:v_imp})).
The left panel is for the standard case ($f_d=1$) while the right one is for the massive case ($f_d=3$).
Our results show that $v_{imp}$ is almost constant for the initial stage, and gradually increases with time (see the solid line).
In addition, the results indicate that the contribution from $v_{esc}$ (the dotted line) dominates over that of $v_m$ (the dashed line) at all the stages.
It is important that both $v_{esc}$ and $v_{m}$ have the same behavior as time goes on.
This occurs, because they are intimately coupled with planetary growth (see Figure \ref{fig1}).
Since planetary growth is regulated predominantly by the dynamics of small planetesimals (see Section \ref{model}),
it is crucial to model $v_m$ accurately to compute $v_{esc}$ (and hence $v_{imp}$),
even if $v_m$ provides a minor contribution to $v_{imp}$.
Note that the low value of $v_m$ is a simple reflection of gas drag that efficiently decreases the eccentricity of small planetesimals.

Also, Figure \ref{fig2} (top) clearly shows that most stages of planetary accretion cannot achieve a high value of $v_{imp}$ 
which can lead to chondrule formation (see the red, solid line).
Only an exception is the final stage where a protoplanet undergoes oligarchic growth and approaches $M_{p,iso}$.
This therefore suggests that chondrule-forming impacts can occur only at the final stage of planetary accretion
for protoplanet-planetesimal collisions.
To confirm this trend rigorously,
it is important to take into account the distribution of $v_m$ (or $e$),
rather than simply considering $\langle e_{eq}^2 \rangle^{1/2}$.
Accordingly, we compute a fractional number of planetesimals ($F_{pl}$)
that can satisfy the condition that $v_{imp} > v_c$ (see the middle panel of Figure \ref{fig2}). 
Let us assume that the equilibrated eccentricity of small planetesimals is represented well by the Rayleigh distribution \citep[e.g.,][]{ki00},
and that when they are colliding with a protoplanet, their impact velocity can be given by equation (\ref{eq:v_imp}).
Then, the value of $F_{pl}(v_{imp} > v_{c})$ can be obtained 
either by numerically integrating the value of $F_{pl}(v_{imp} > v_c)$ 
or analytically computing the value of 
\begin{eqnarray}
\label{eq:F_pl}
F_{pl}(v_{imp}>v_c) & \equiv & Ra(v_m^2 > v_c^2 - v_{esc}^2)  \\ \nonumber
                                & \simeq & \exp \left( - \frac{ e_c^2 - (v_{esc}/v_{Kep})^2} { \langle e_{eq}^2 \rangle } \right)  \\ \nonumber
                                & =          & \exp \left( - \frac{v_c^2 - v_{esc}^2 }{  v^2_{m} }  \right),
\end{eqnarray} 
where $Ra(v_m)$ is a function to represent the Rayleigh distribution.
Note that $v_m = \langle e_{eq}^2 \rangle^{1/2}_{run} v_{Kep}$ when a protoplanet undergoes runaway growth,
while  $v_m = \langle e_{eq}^2 \rangle^{1/2}_{oli} v_{Kep}$ for oligarchic growth.
Figure \ref{fig2} (middle) shows that the resultant value of $F_{pl} (v_{imp} > v_c)$ is almost zero, except for a certain time
when $v_{imp}$ exceeds $v_c$ (see the hatched region).
We define $\tau=\tau_{int}$ such that $v_{imp}= v_c$.
The value of $F_{pl}(v_{imp}> v_c)$ becomes unity until oligarchic growth is complete (i.e., $\tau=\tau_{end}$).
Thus, we demonstrate that protoplanet-planetesinal collisions can result in chondrule formation
only if a protoplanet undergoes oligarchic growth and becomes massive enough,
so that colliding small planetesimals obtain the value of $v_{imp}$ that is higher than $v_c$.

Provided that the value of $F_{pl}(v_{imp}>v_c)$ is given,
we can finally compute the cumulative mass of chondrules ($\Delta m_{ch,M-m}$) that are formed by impact jetting of protoplanet-planetesimal collisions.
This can be done by numerically integrating the following equation;
\begin{equation}
 \label{eq:dm_M-m}
 \frac{d \Delta m_{ch,M-m}}{dt} \simeq \frac{dM_p}{dt} F_{ch} F_{pl}(v_{imp} > v_c) ,
\end{equation}
where $d M_p/dt$ is given by equation (\ref{eq:dmdt}) and 
$F_{ch} =10^{-2}$ is a fractional mass of planetesimals that can eventually generate chondrules by impact jetting,
following \citet{jmm15} (see Table \ref{table1}).
In equation (\ref{eq:dm_M-m}), the factor of $(dM_p/dt)F_{ch}$ represents how much mass of chondrules can be created
as a protoplanet grows,
and the factor, $F_{pl}(v_{imp}>v_c)$, denotes how many collisions can satisfy the chondrule formation condition that $v_{imp}>v_c$.

We present the time evolution of the resultant $\Delta m_{ch,M-m}$ in Figure \ref{fig2} (bottom).
The results show that the value of $\Delta m_{ch,M-m}$ increases rapidly with time
only for $\tau_{int} < \tau < \tau_{end}$ (see the hatched region, also see Table \ref{table1}).
This is simply because $F_{pl}$ becomes unity at that time.
Once a protoplanet obtains the isolation mass ($M_{p,iso}$) and planetary accretion is complete (i.e., $\tau=\tau_{end}$),
the value of $\Delta m_{ch,M-m}$ becomes constant.
It is of fundamental importance that 
this saturated value of $\Delta m_{ch,M-m}$ is essentially comparable to that of $F_{ch} M_{p,iso}$ (see the dotted line).
This arises because chondrule-forming impacts occur when a protoplanet accretes most of its mass 
(i.e., $\tau_{int} < \tau < \tau_{end}$, see Figure \ref{fig1}).
Thus, we can conclude that the total mass of chondrules formed by protoplanet-planetesimal collisions can 
readily be estimated if growing protoplanets reach the value of $M_{p,iso}$ well before disk lifetimes.

For the massive disk case ($f_d=3$), 
all the results are qualitatively similar to those for the standard case (see the right panel in Figure \ref{fig2}).
As pointed out in Section \ref{model}, only a slight difference exists;
massive disks can trigger chondrule-forming collisions at much earlier stages (see the hatched region),
since such disks accelerate planetary accretion.
It is important to emphasize that the relationship that $\Delta m_{ch,M-m} \simeq F_{ch} M_{p,iso}$ 
is still maintained for this case.

\subsection{Planetesimal-planetesimal collisions} \label{res_pl}

We then consider planetesimal-planetesimal collisions and discuss how such collisions can contribute to chondrule formation.
As done in Section \ref{res_iso}, we compute the impact velocity ($v_{imp}$), a fractional number of planetesimals ($F_{pl}(v_{imp} > v_c)$), 
and the cumulative mass of chondrules ($\Delta m_{ch,m-m}$).
Note that $\Delta m_{ch,m-m}$ is newly defined to differentiate the previous results ($\Delta m_{ch,M-m}$).
We consider planetary accretion at $a=1$ AU again.

\begin{figure*}
\begin{minipage}{17cm}
\begin{center}
\includegraphics[width=8cm]{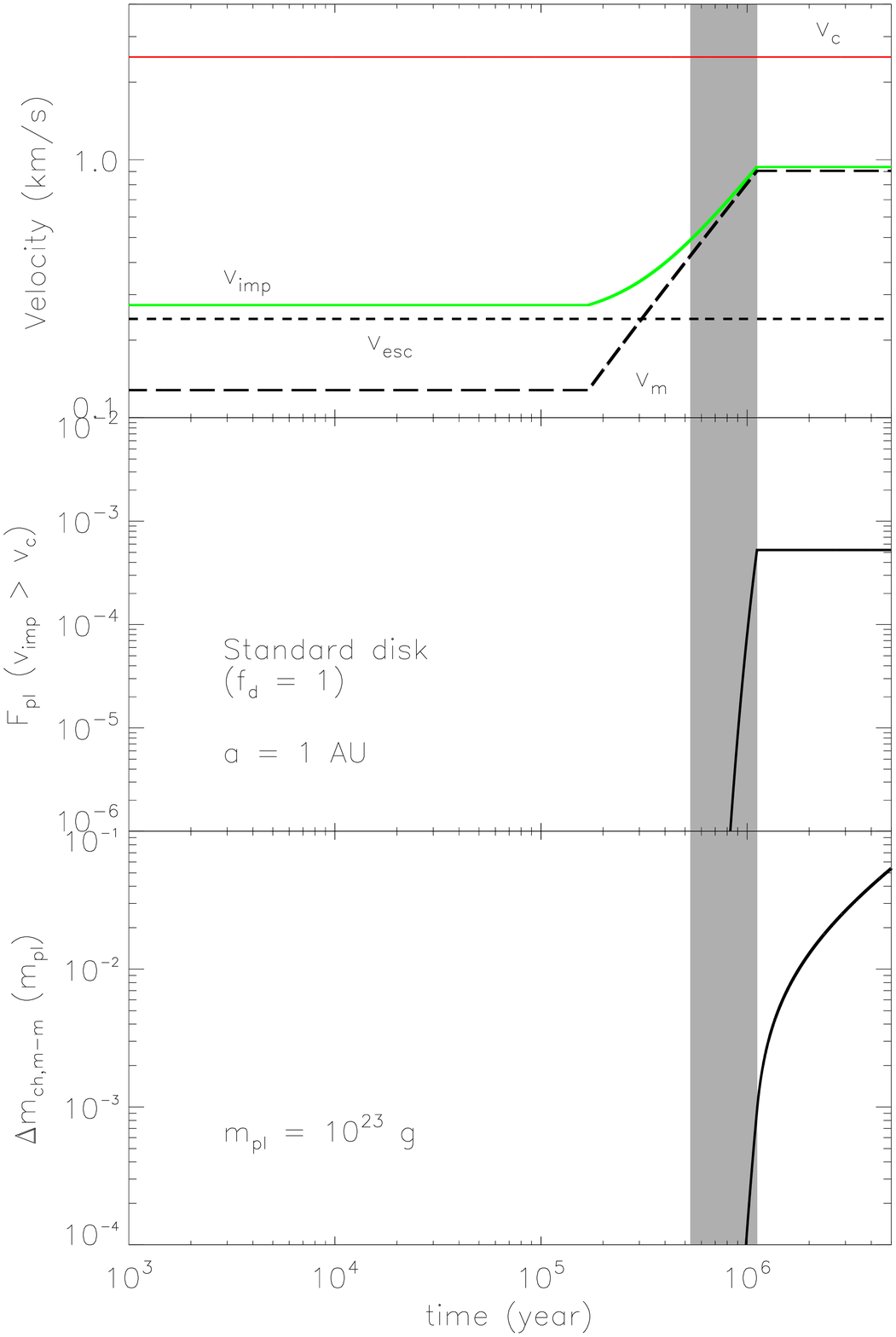}
\includegraphics[width=8cm]{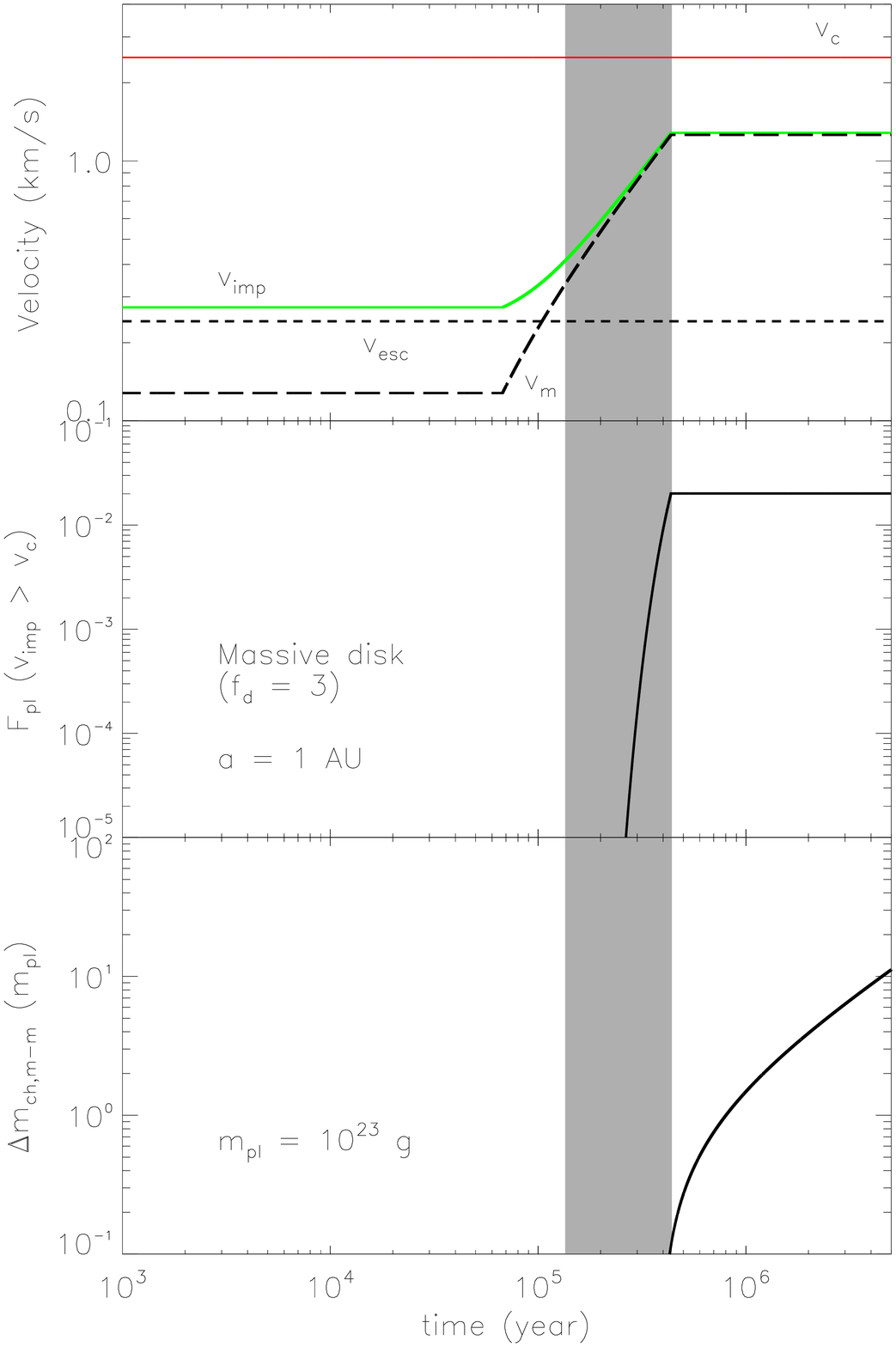}
\caption{The computed value of the impact velocity ($v_{imp}$), $F_{pl}(v_{imp} > v_c)$, and the cumulative chondrule mass ($\Delta m_{ch,M-m}$) 
as a function of time on the top, the middle, and the bottom, respectively, at the position of $a=1$ AU (as Figure \ref{fig2}).
Planetesimal-planetesimal collisions are considered.
To visualize the difference with the results of protoplanet-planetesimal collisions,
the hatched region is exactly identical to that in Figure \ref{fig2}.
The results show that planetesimal-planetesimal collisions can contribute to the mass budget of chondrules 
once oligarchic growth is complete, but its effect is very minor.}
\label{fig3}
\end{center}
\end{minipage}
\end{figure*}

Figure \ref{fig3} (top) shows the resultant behavior of $v_{imp}$ (as in Figure \ref{fig2}).
The results show that $v_{imp}$ is initially constant and increases with time as oligarchic growth proceeds.
There are two main differences with the protoplanet-planetesimal collisions.
First, even at the later stage of oligarchic growth (see the hatched region),
$v_{imp}$ cannot reach the value of $v_c$.
This originates from $v_m$ that can provide a dominant contribution to $v_{imp}$ at that time
for planetesimal-planetesimal collisions.
As discussed in Section \ref{model}, field planetesimals do not grow very much during both runaway and oligarchic growth.
Consequently, $v_{esc}$ remains small and roughly constant for all the stages,
which can eventually be overcome by $v_{m}$.
Second, $v_{imp}$ can maintain the constant value even after planetary accretion is complete.
This is simply because planetesimal-planetesimal collisions are considered here;
even if a protoplanet obtains the isolation mass and consumes all the planetesimals in its feeding zone,
(which results effectively in no collision with a protoplanet),
field planetesimals outside the feeding zone can continue to collide with other planetesimals.
Thus, when planetesimal-planetesimal collisions are considered,
most of potentially chondrule-forming impacts can occur after a protoplanet obtains $M_{p,iso}$ (i.e., $\tau>\tau_{end}$).

We examine this possibility further by computing $F_{pl} (v_{imp} > v_c)$.
To proceed, we again assume the Rayleigh distribution for the equilibrium eccentricity of small planetesimals 
($v_m$, see equation (\ref{eq:F_pl})).
Since planetesimal-planetesimal collisions are considered, 
we numerically perform the time-integration of the following equation;
\begin{equation}
\label{eq:dm_M-m2}
\frac{ d \Delta m_{ch,m-m} }{dt} \simeq  [ \pi \Sigma_d a^2 k_{col} ] F_{ch}  F_{pl} (v_{imp} > v_c),
\end{equation}
where $k_{col}$ is the collision rate between planetesimals and is given as \citep[e.g.,][]{im93}
\begin{equation}
 \label{eq:k_col}
 k_{col} \equiv \frac{1}{m_{pl}} \frac{d m_{pl}}{dt} \simeq  \pi \Sigma_d \frac{ G r_{pl}}{\langle e_{eq}^2 \rangle a v_{Kep}}.
\end{equation}
Note that $\langle e_{eq}^2 \rangle = \langle e_{eq}^2 \rangle_{run}$ when a protoplanet undergoes runaway accretion,
while $\langle e_{eq}^2 \rangle = \langle e_{eq}^2 \rangle_{oli}$ for oligarchic growth.
In equation ($\ref{eq:dm_M-m2}$), we have assumed that $ 2 \Delta a \simeq a$,
so that the factor, $ [ \pi \Sigma_d a^2 k_{col} ] F_{ch}$, can denote how much mass of chondrules can be generated 
via planetesimal-planetesimal collisions at $a=1$ AU for a certain time.
The factor, $F_{pl} (v_{imp} > v_c)$, can correspond to the total number of collisions 
that can meet the formation condition of chondrules.

Figure \ref{fig3} (middle) shows its outcome;
it is obvious that chondrule-forming impacts can be established 
even after the formation of a protoproplanet is finished (so $\tau>\tau_{end}$).
The results indicate that this stage gives the largest value of $F_{pl} (v_{imp} > v_c)$,
since viscous stirring of a protoplanet acting on small planetesimals is most significant.
The maximum value of $F_{pl}(v_{imp}>v_c)$ is however much lower than that for protoplanet-planetesimal collisions.
Thus, we demonstrate that planetesimal-planetesimal collisions can provide only a minor contribution to chondrule formation,
as long as forming protoplanets obtain $M_{iso}$ within disk lifetimes (i.e., $\tau_{end} < \tau_{d}$).
In other words, the difference in the onset of chondrule formation 
between protoplanet-planetesimal and planetesimal-planetesimal collisions 
may not be so crucial, 
especially at $a \la 2-3$ AU (see the following discussion and Figure \ref{fig4}).
Figure \ref{fig3} (bottom) shows the cumulative mass of chondrules ($\Delta m_{ch,m-m}$) as a function of time,
and confirms the above arguments.  

When massive disks are considered (see the right panel of Figure \ref{fig3}),
the qualitatively similar results are obtained,
while chondrule formation begins at earlier stages and the resultant value of $\Delta m_{ch,m-m}$ becomes higher.
This is simply because the value of $M_{p,iso}$ is larger for this case (see Figure \ref{fig1}),
which increases both $v_{imp}$ and $F_{pl}(v_{imp} > v_c)$, and shortens $\tau_{int}$ and $\tau_{end}$.

\subsection{Timescales of chondrule formation} \label{res_t}

As discussed in Sections \ref{res_iso} and \ref{res_pl},
the formation of protoplanets and the resultant protoplanet-planetesimal collisions can act as the major source of chondrule formation.
In this section, we specify the range of $a$ at which the formation of protoplanets can complete within disk lifetimes,
and hence identify when chondrule formation occurs by protoplanet-planetesimal collisions.

In principal, we can follow the time evolution of planetary accretion and 
compute both $\tau_{int}$ and $\tau_{end}$ with $a$ varying.
For this case, $\tau_{int}$ is defined so that $v_{imp} = v_c$,
and $\tau_{end}$ is obtained when oligarchic growth is complete (as done in Figure \ref{fig2}).
Here, we attempt to carry out fully analytical calculations to estimate the value of $\tau_{int}$ and $\tau_{end}$.
This can be done as what follows;
for $\tau_{int}$, we first assess the mass of protoplanets that satisfy the condition that $v_{imp} = v_c$,
by assuming that 
\begin{equation}
v_{imp} \approx v_{esc} \approx \sqrt{ \frac{2G M_p}{R_p} }.
\end{equation}
Then, the resultant value of $M_p$ is used to compute $\tau_{int}$,
which can be given as
\begin{equation}
\tau_{int} \equiv f_{\tau} \frac{M_p}{ dM_p /dt},
\end{equation}
where oligarchic growth is assumed in calculating the value of $dM_p/dt$ (see equation (\ref{eq:dmdt})).
Note that a factor, $f_{\tau}$, arises from consideration that 
an actual time ($\tau_{int}$) that is needed for growing protoplanets to meet the condition that $v_{imp} = v_c$,
can be scaled by the growth timescale ($M_p / (d M_p/dt)$).
In fact, we find that fully analytical calculations can well reproduce the results of time-dependent ones
when $f_{\tau} =3$ for most cases of our calculations (see below). 
For $\tau_{end}$, calculations become straightforward as 
\begin{equation}
\tau_{end} \equiv  \left. f_{\tau} \frac{M_{p}}{ dM_{p} /dt } \right|  _{M_p = M_{p,iso}},
\end{equation} 
where equation (\ref{eq:mp_iso}) is utilized for computing $M_{p,iso}$, and 
a factor $f_{\tau}=3$ is introduced again.

\begin{figure*}
\begin{minipage}{17cm}
\begin{center}
\includegraphics[width=8cm]{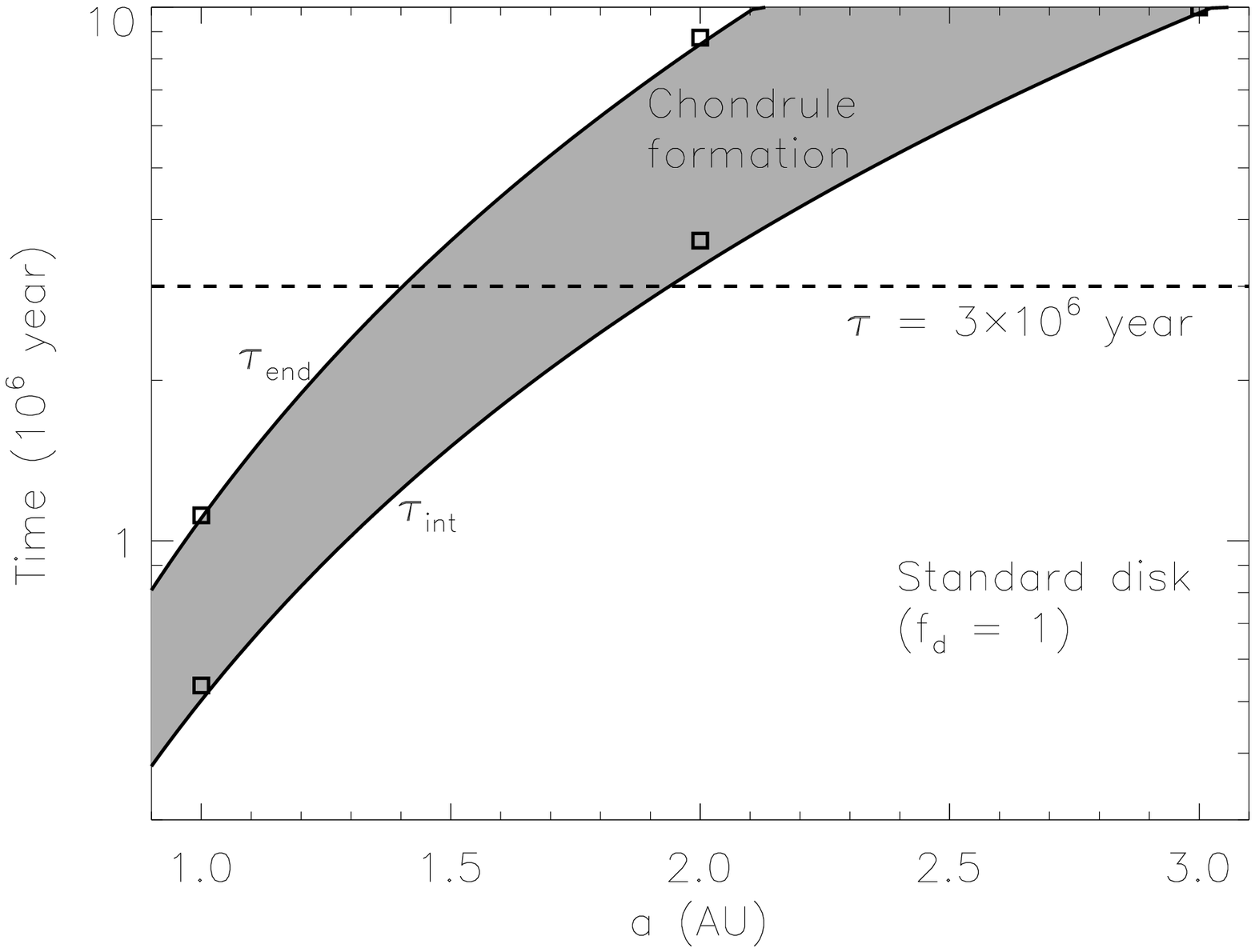}
\includegraphics[width=8cm]{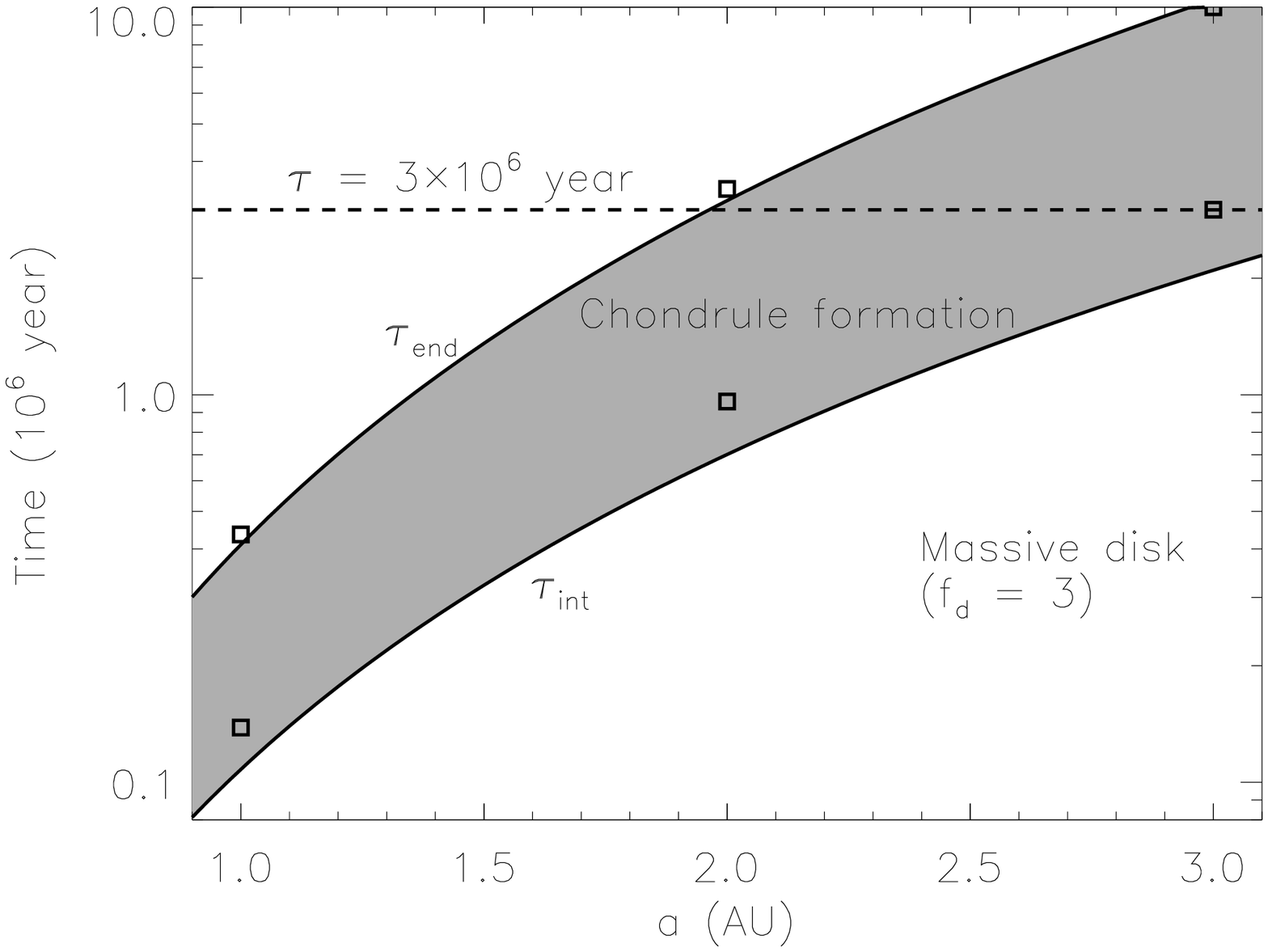}
\caption{The value of $\tau_{int}$ and $\tau_{end}$ as a function of $a$.
While the standard case ($f_d =1$) is shown on the left panel,
the massive case ($f_d =3$) is on the right one.
For comparison purpose, the results of the time-dependent calculations are denoted by the squares.
Following planetary accretion that becomes slower as $a$ increases,
both $\tau_{int}$ and $\tau_{end}$ become an increasing function of $a$.
The value of $\tau_{end}$ approaches at $\tau=10^7$ years at $a=2$ AU for the standard case (the left panel), 
and at $a=3$ AU for the massive case (the right panel).
Since $\tau_d = 10^7$ years may be an upper limit of disk lifetimes, 
chondrule formation via impact jetting can occur within these orbital radii (see the hatched region).}
\label{fig4}
\end{center}
\end{minipage}
\end{figure*}

Figure \ref{fig4} shows the resultant $\tau_{int}$ and $\tau_{end}$ as a function of $a$.
The standard case ($f_d=1$) is presented on the left panel,
and the massive one ($f_d=3$) is on the right one.
The results of time-dependent calculations are also plotted 
to confirm how valid fully analytical calculations are (see the squares). 
As expected, the value of both $\tau_{int}$ and $\tau_{end}$ increases steadily with $a$,
(since planetary accretion becomes slower for larger $a$).
The results show that for the standard case ($f_d=1$), $\tau_{end}$ approaches $\tau=10^7$ years at $a=2$ AU,
while that occurs at $a=3$ AU for the massive case ($f_d=3$).
It is currently recognized widely that an upper limit of gas disk lifetimes is about $\tau_d \simeq 10^7$ years \citep[e.g.,][]{wc11}.
In the following discussion, therefore, we consider a certain range of $a$;
for standard disks, $0.1\mbox{ AU} < a < 2 \mbox{ AU}$,
and for massive disks, $0.1\mbox{ AU} < a < 3 \mbox{ AU}$.
Note that these ranges may be reasonable for chondrule formation,
because chondrule formation may not take place far beyond the water-ice line 
that is located at $a = 3-4$ AU \citep[e.g.,][]{cc06i}.

\subsection{The $a-$dependence} \label{res_a}

We are now in a position to compute the total mass of chondrules ($\Delta m_{ch}$) that are formed by impact jetting 
via both protoplanet-planetesimal and planetesimal-planetesimal collisions.
We here  examine how $\Delta m_{ch,M-m}$, $\Delta m_{ch,m-m}$, and the resultant $\Delta m_{ch}$ behave 
as a function of the distance ($a$) from the central star.

Since we have confirmed that 
protoplanet-planetesimal collisions can complete within $\tau=\tau_d$ for a certain range of $a$ (see Figure \ref{fig4}),
we here adopt the simplified, time-independent calculations.
Then, $\Delta m_{ch}$ can be given as
\begin{equation}
 \label{eq:m_ch}
 \Delta m_{ch} \equiv \Delta m_{ch,M-m} + \Delta m_{ch,m-m}
\end{equation}
where
\begin{equation}
 \label{eq:m_ch_mpl_m}
\Delta m_{ch,M-m} \simeq F_{ch} M_{p,iso},
\end{equation}
for protoplanet-planetesimal collisions (see Figure \ref{fig2}), and for planetesimal-planetesimal ones,
\begin{equation}
\label{eq:m_ch_m_m}
\Delta m_{ch,m-m} \simeq [ \pi \Sigma_d a^2 F_{pl} (v_{imp} > v_c) ]\times F_{ch} \times ( k_{col} \tau_d ),
\end{equation}
where we have assumed that $2 \Delta a \simeq a$ again.
In this equation, the first bracket in the right hand side represents the total mass of planetesimals 
that satisfy $v_{imp} > v_c$ at the orbital distance of $a$ at a certain time, 
the second one does the fractional mass of a planetesimal 
that can eventually be transformed into chondrules via impact jetting through one collision,
and the last one does the total number of collisions over the disk lifetime.
Note that our estimate of $\Delta m_{ch,m-m}$ would be an upper limit, 
since we assume that $M_p = M_{p,iso}$ in calculating both $F_{pl} (v_{imp} > v_c)$ and $ k_{col}$,
and that a considerable number of field planetesimals can exist even after $M_p = M_{p,iso}$.

\begin{figure*}
\begin{minipage}{17cm}
\begin{center}
\includegraphics[width=8cm]{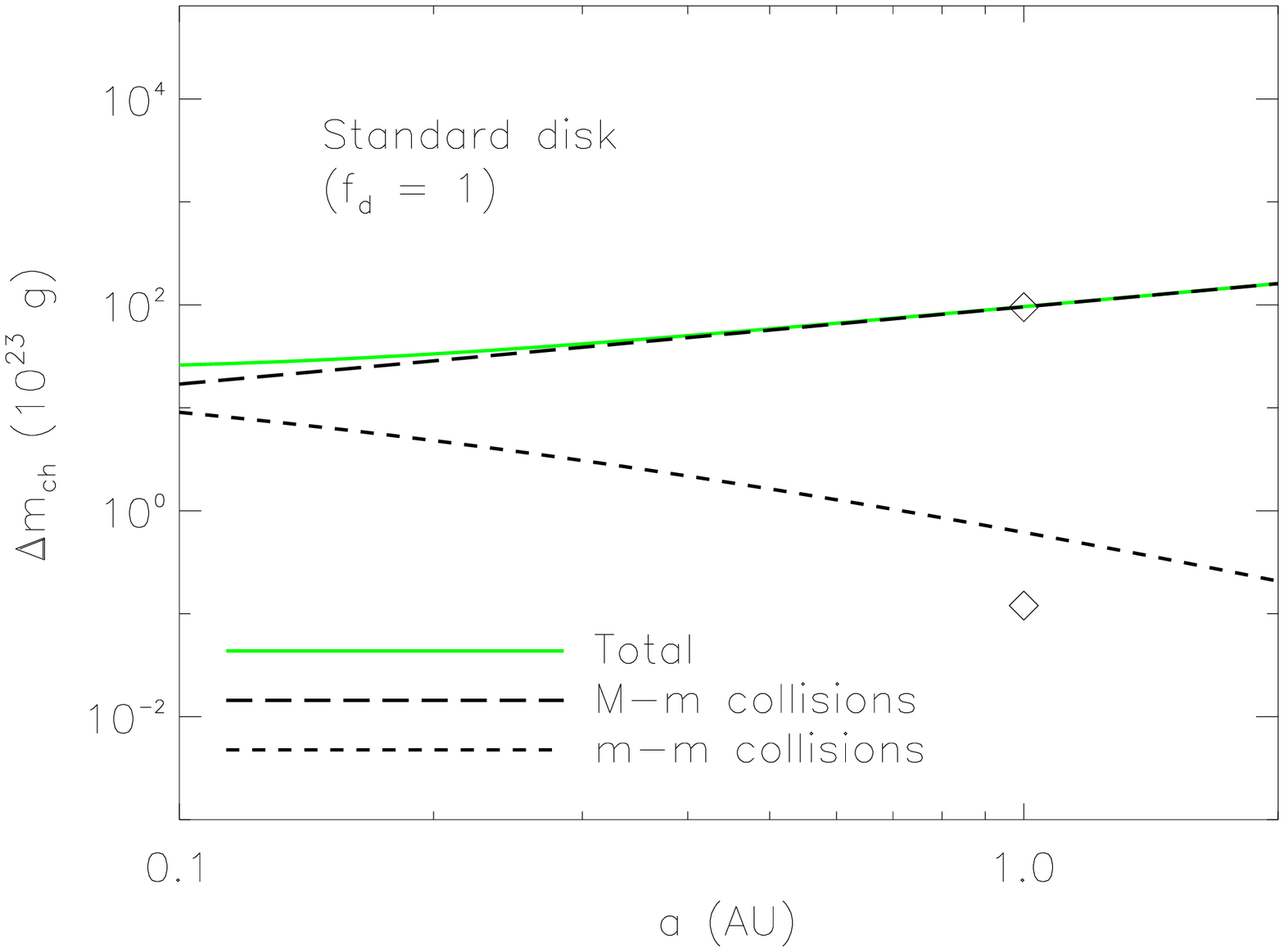}
\includegraphics[width=8cm]{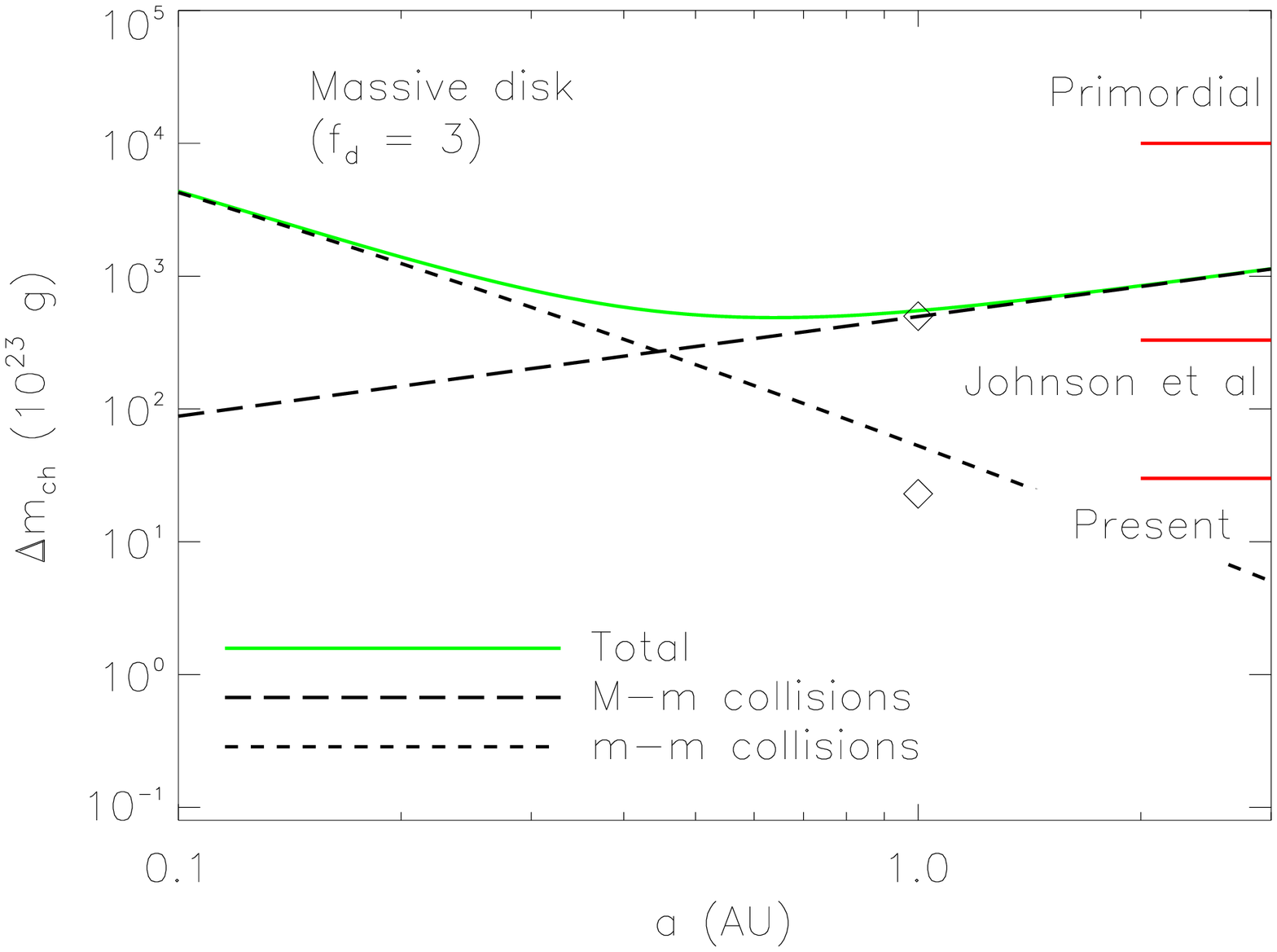}
\caption{The total mass of chondrules ($\Delta m_{ch}$) formed by impact jetting as a function of  the distance ($a$) from the central star.
The standard case ($f_d=1$) is shown on the left panel and the massive case ($f_d=3$) is on the right one.
The total value ($\Delta m_{ch}$) is denoted by the green, solid line (see equation (\ref{eq:m_ch})),
while that of $\Delta m_{ch,M-m}$ and $\Delta m_{ch,n-m}$ is 
by the dashed (see equation (\ref{eq:m_ch_mpl_m})) and the dotted line (see equation (\ref{eq:m_ch_m_m})), respectively.
In order to compare the time-independent results with those derived from the time-dependent approach (see Sections \ref{res_iso} and \ref{res_pl}), 
the latter ones are also plotted (see the diamonds). 
The results show that $\Delta m_{ch,M-m}$ provides a major contribution to $\Delta m_{ch}$ 
especially in the outer part of disks for a given range of $a$.
As a reference, we also plot the value of the primordial asteroid belt ($\Delta m_{ch} \simeq 10^{27}$ g),
the results of \citet{jmm15} ($\Delta m_{ch} \simeq 3.3 \times 10^{25}$ g),
and the value of the current asteroid belt ($\Delta m_{ch} \simeq 3 \times 10^{24}$ g).}
\label{fig5}
\end{center}
\end{minipage}
\end{figure*}

Figure \ref{fig5} shows the results of $\Delta m_{ch}$.
While the standard case ($f_d=1$) is presented on the left panel,
the massive case ($f_d=3$) is on the right one.
In order to verify our simplified approach, 
(see equations (\ref{eq:m_ch}), (\ref{eq:m_ch_mpl_m}), and (\ref{eq:m_ch_m_m})),
we also plot the value of $\Delta m_{ch,M-m}$ and $\Delta m_{ch,m-m}$ 
that are obtained in Sections \ref{res_iso} and \ref{res_pl}, respectively (see the diamonds).
The results confirm that the time-independent approach leads to a reasonable agreement with the time-dependent calculations.
As anticipated, the contribution of $\Delta m_{ch,M-m}$ (see the dashed line) dominates over that of $\Delta m_{ch,m-m}$ (see the dotted line),
and hence the value of $\Delta m_{ch}$ (see the green, solid line) is determined largely by $\Delta m_{ch,M-m}$.
It is interesting that the opposite situation is realized in the inner part of disks ($a \la 0.4$ AU) 
when the massive case is adopted (see the right panel).
This occurs because $M_{p,iso} \propto a^{3}$ while $k_{col}$ roughly goes as $a^{-3}$,
which ends up with $\Delta m_{ch,m-m}$ that becomes a decreasing function of $a$.

Thus, our results indicate that protoplanet-planetesimal collisions play a crucial role in forming chondrules by impact jetting
in the region of the present asteroid belt, when disk lifetimes are long enough ($\tau_d \simeq 10^7$ years).

\subsection{The $m_{pl}-$dependence}

We have so far assumed that $m_{pl} = 10^{23}$ g
and neglected the size distribution of planetesimals.
Here we explore its effect, by simply changing the value of $m_{pl}$.
As an example, we consider the case that $m_{pl} = 10^{24}$ g and $f_d=3$, 
and carry out the same calculations as above.

\begin{figure*}
\begin{minipage}{17cm}
\begin{center}
\includegraphics[width=8cm]{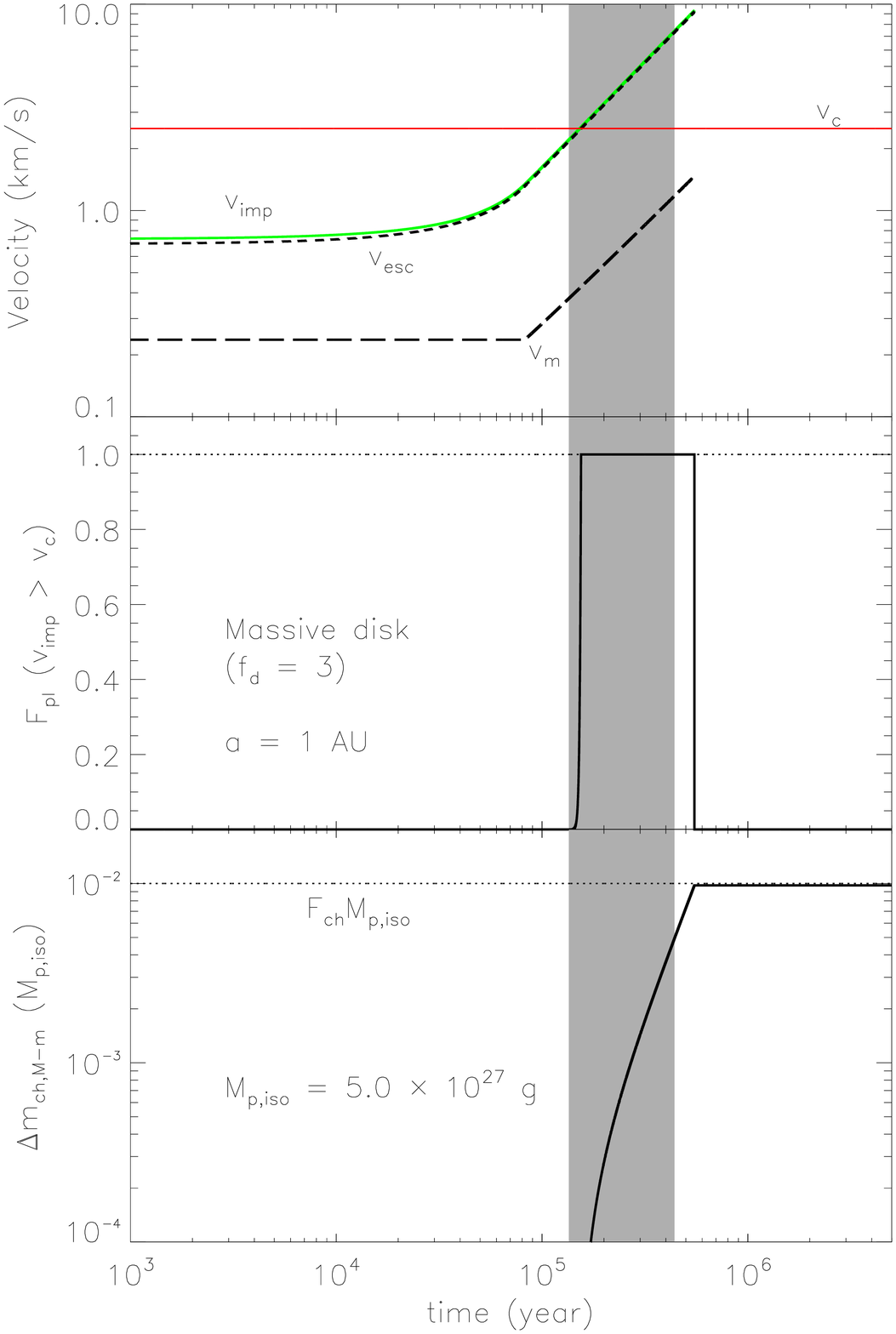}
\includegraphics[width=8cm]{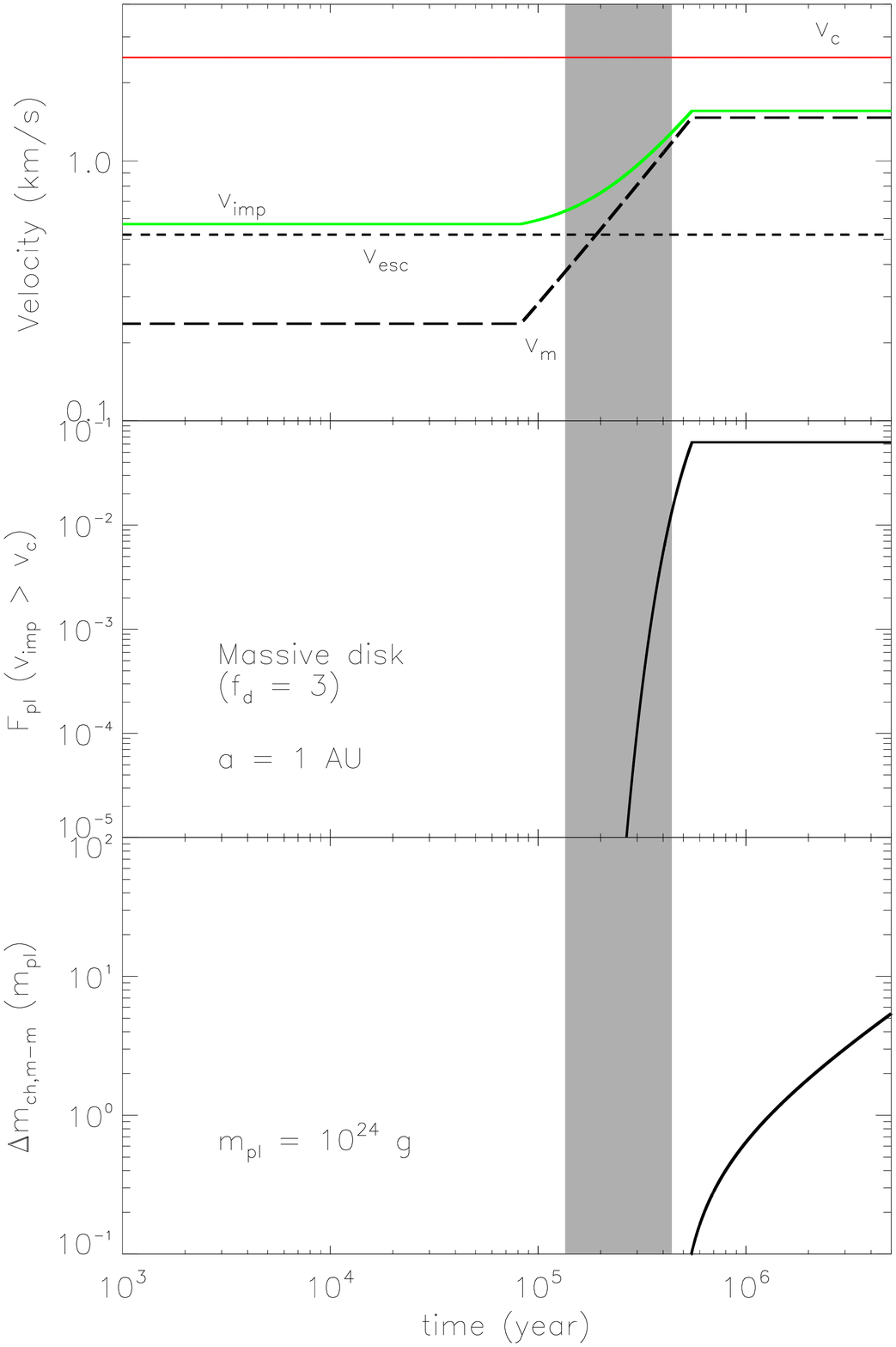}
\caption{The results of $v_{imp}$, $F_{pl}(v_{imp}>v_c)$, and $\Delta m_{ch}$ as a function of time 
for the case that $m_{pl} = 10^{24}$ g and $f_d=3$ (as Figures \ref{fig2} and \ref{fig3}).
While the results of protoplanet-planetesimal collisions are shown on the left panel, 
those of planetesimal-planetesimal collisions are on the right panel.
When massive planetesimals are considered, 
planetary accretion triggers chondrule-forming collisions at slightly later stages,
which is caused by the slower growth rate of a protoplanet.
This feature is visualized by plotting the hatched region that is obtained in Figures \ref{fig2} and \ref{fig3} (see their right panel).}
\label{fig6}
\end{center}
\end{minipage}
\end{figure*}

\begin{figure}
\begin{center}
\includegraphics[width=8cm]{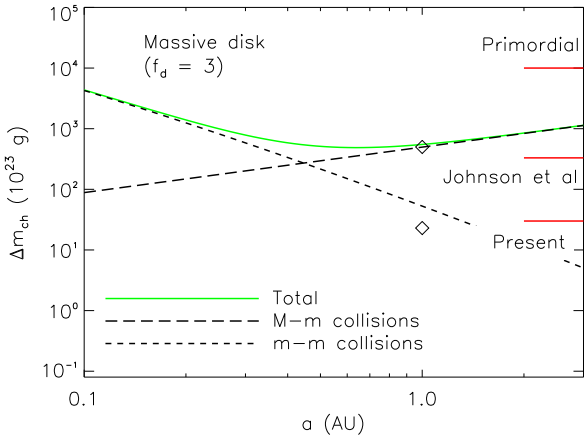}
\caption{The value of $\Delta m_{ch}$, $\Delta m_{ch,M-m}$, and $\Delta m_{ch,m-m}$ as a function of $a$
for the case that $m_{pl}=10^{24}$ g and $f_d=3$ (as Figure \ref{fig5}).
The results verify that $\Delta m_{ch,M-m}$ plays a major role in estimating $\Delta m_{ch}$ in the region of the present asteroid belt 
even if the value of $m_{pl}$ varies.}
\label{fig7}
\end{center}
\end{figure}

Figures \ref{fig6} and \ref{fig7} show the corresponding results;
the time-dependent calculations of $v_{imp}$, $F_{pl}(v_{imp}>v_c)$, $\Delta m_{ch,M-m}$, and $\Delta m_{ch,m-m}$ 
are presented in Figure \ref{fig6} (as in Figures \ref{fig2} and \ref{fig3}),
and the time-independent calculations of $\Delta m_{ch}$, $\Delta m_{ch,M-m}$, and $\Delta m_{ch,m-m}$ are on Figure \ref{fig7} (as in Figure \ref{fig5}).
We find that the increase of $m_{pl}$ delays the onset of chondrule-forming impacts (see Figure \ref{fig6}).
This is a consequence of planetary accretion that also gets delayed; 
when massive planetesimals serve as the main agent to collide with a growing protoplanet,
the eccentricity damping caused by gas drag becomes less efficient (see equation (\ref{eq:e_oli})),
and hence planetary growth slows down (see equation (\ref{eq:dmdt})).
To illustrate this feature clearly in Figure \ref{fig6}, 
we plot the hatched region that is exactly identical to that on the right panel of Figure \ref{fig2}.
It is important that we again obtain that $\Delta m_{ch,M-m} \simeq F_{ch} M_{p,iso}$ and 
$\Delta m_{ch,M-m} \gg \Delta m_{ch,m-m}$,
while $\Delta m_{ch,m-m}$ is an increasing function of $m_{pl}$.
This trend is indeed confirmed in Figure \ref{fig7}.

\begin{figure}
\begin{center}
\includegraphics[width=8cm]{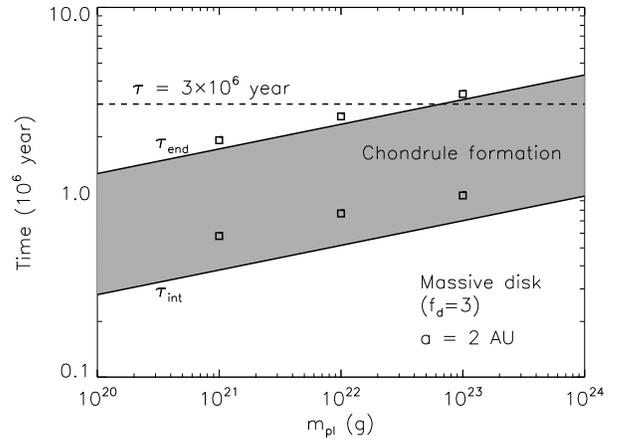}
\caption{The results of $\tau_{int}$ and $\tau_{end}$ as a function of $m_{pl}$ (as Figure \ref{fig4}).
As an example, massive disks ($f_d=3$) and the position of $a=2$ AU are considered.
The results show that both $\tau_{int}$ and $\tau_{end}$ are much shorter than the disk lifetime ($\tau_d = 10^7$ years)
for a wide range of $m_{pl}$.
This confirms that oligarchic growth there can complete within the disk lifetime.}
\label{fig8}
\end{center}
\end{figure}

We also investigate the effect of $m_{pl}$ on the value of $\tau_{int}$ and $\tau_{end}$.
This is important because protoplanet-planetesimal collisions contribute predominantly to the mass budget of chondrules,
so that we should examine whether or not such collisions can complete within disk lifetimes ($\tau_{d}$),
even if the value of $m_{pl}$ is changed.
Following the same procedure as done in Section \ref{res_t},
we compute $\tau_{int}$ and $\tau_{end}$ as a function of $m_{pl}$ at $a=2$ AU (see Figure \ref{fig8}).
The results show that the value of $\tau_{end}$ is smaller than $\tau_d \simeq 10^7$ years for all the values of $m_{pl}$.
Thus, we can conclude that protoplanet-planetesimal collisions 
can be regarded as a major process to form chondrules via impact jetting for a wide range of $m_{pl}$, 
and our time-independent calculations that are done in Figure \ref{fig7} are reasonable.

In summary, our results demonstrate that even if the size distribution of $m_{pl}$ is taken into account,
the overall feature of chondrule formation by impact jetting does not change very much.
 
\section{Discussion} \label{disc} 

We discuss how important planetary accretion is to form chondrules 
by comparing our results both with values that are derived from the currently available samples, 
and with the results of \citet{jmm15}.
We also explore the effect of other physical processes that can potentially affect our results, 
and discuss their possible outcomes.

\subsection{Implications for chondrule formation}

We attempt to derive a definitive remark of the significance of planetary accretion for forming chondrules.

To this end, we compare the results of $\Delta m_{ch}$ with two characteristic values: 
the one is the currently measured mass of the present asteroid belt, which is $\simeq 3 \times 10^{24}$ g \citep[e.g.,][]{dc14}.
The other is the computed value by numerically tracing the collisional history of the main asteroid belt,
which is $\simeq 10^{27}$ g \citep[e.g.,][]{bdn05}.
The former one may provide a lower limit.
Note that, while it is not entirely obvious what fraction of the total mass should be occupied by chondrules in the present asteroid belt,
an order-of-magnitude estimate suggests that the total mass of chondrules in the primordial asteroid belt 
may be a few times that of the current asteroid beld \citep[e.g.,][]{hca09}.
The latter one may be an upper limit and can be regarded as the total mass of the primordial asteroid belt.
The value itself may be prevalent for the asteroid community \citep[e.g.,][]{md09}.

Figure \ref{fig5} (the left) shows clearly that 
our results of $\Delta m_{ch}$ are located between the upper and the lower limits (also see Figure \ref{fig7}).
More specifically, the results indicate that, when massive disks ($f_d=3$) are considered,
the resultant value becomes an order of $10^{26}$ g that is about 1 \% of the isolation mass there,
and it is independent of the value of $m_{pl}$.
As found by \citet{jmm15}, therefore, 
impact jetting triggered by planetary accretion is an intriguing process to understand an origin of chondrules.
More dedicated work would be needed to fully validate whether or not the resultant value of $\Delta m_{ch}$ is good enough to 
reproduce the current abundance of chondrules in chondrites.

How about the timescale of chondrule formation by impact jetting?
It is well recognized that chondrule formation occurred in the solar nebula 
about 0 $\sim $ 3 million years after CAI formation (Section \ref{intro}).
As discussed above, 
our results indicate that the formation timing of chondrules ($\tau_{end}$ and $\tau_{end}$) depends 
both on the orbital distance ($a$) (see Figure \ref{fig4}) and on the mass of planetesimals ($m_{pl}$) (see Figure \ref{fig8}).
Figure \ref{fig4} (the left) shows that, for massive disks, 
the majority of chondrules can form within $\tau \simeq 3 \times 10^6$ year at $a=2$ AU,
and within $\tau \simeq 10^7$ years at $a=3$ AU.
Figure \ref{fig8} depicts that the formation time of chondrules can be shorter
when planetesimals with $m_{pl}< 10^{22}$ g involve chondrule-forming impacts.
Thus, impact jetting is promising in the sense that, for a reasonable range of $a$ and $m_{pl}$,
it can lead to a timescale that is roughly consistent with the necessary condition that is measured for chondrules in chondrites  \citep[e.g.,][]{cbk12}.
This is also pointed out by  \citet{jmm15}.
It should be noted clearly that 
we do not explicitly define when planetesimal formation begins,
so that $\tau=0$ in our calculations is not necessarily identical to the time when CAI formation began.

As a conclusion, our results demonstrate that, based on a current picture of planetary accretion, 
impact jetting triggered by planetesimal collisions can act as an important process 
to reproduce a couple of chondrules' properties that are derived from the current samples of chondrites.

\subsection{Comparison with \citet{jmm15}}

As demonstrated above, our results essentially lead to the same conclusions as those of \citet{jmm15},
even if we have adopted a much simpler approach.
This originates simply from a consequence 
that the resultant value of $\Delta m_{ch}$ is determined predominantly by the value of $M_{p,iso}$.
In other words, it is crucial to complete the formation of protoplanets for a given disk lifetime.
As shown in Figure \ref{fig4} (the left),
planetary accretion can be completed within $\tau \simeq 3-10 \times 10^6$ years at $a=2-3$ AU 
for massive disks ($f_d =3$).
In addition, the results do not depend on the value of $m_{pl}$ very much (see Figure \ref{fig8}).
Thus, it is very reasonable that both our and their results are comparable.

\subsection{Effects of other processes}

As discussed above, the abundance of chondrules formed by impact jetting does not vary significantly 
under the condition that planetary accretion is completed within a given disk lifetime.
It, however, should be pointed out clearly that the conclusion is valid only if dynamical processes are negligible,
such as gas-induced planetary migration and radial drift acting on the resultant mm-sized chondrules.
In addition, the timescale of chondrule formation can alter due to other physical processes such as disk turbulence and 
perturbation arising from nearby planets such as Jupiter.
We discuss potential effects of these processes on our results.

It is currently recognized well that, when protoplanets form in gas disks,
they experience the so-called planetary migration 
that arises from tidal, resonant interactions between protoplanets and their gas disks \citep[e.g.,][]{kn12,bcp14}.
While the idea of migration was proposed in 1980's \citep[e.g.,][]{gt80},
the detailed behavior of migration, especially its direction, is still a matter of debate.
This is because the recent analytical and numerical studies show that 
the direction of migration is intimately coupled with disk properties 
such as the surface density and the disk temperature \citep[e.g.,][]{pm06,bm08,hp10,kl12,bml14}.
It is therefore non-trivial to quantitatively calibrate the effect of planetary migration 
on the formation of protoplanets and hence on chondrule formation.
Nonetheless, it may be useful to develop a simple, qualitative argument here.
To this end, we follow an approach discussed in \citet{hp12}, 
wherein the timescale of (type I) planetary migration is compared with the local viscous timescale of gas disks (see their equation (27)).
Adopting characteristic values of disk parameters,
we find that planetary migration can play an important role in determining the orbital evolution of protoplanets,
when their mass is larger than $\sim 0.1-1 M_{\oplus}$ at $a\simeq 1-3$ AU (also see Figure 10 of \citet{hp12}).
In other words, the maximum mass of protoplanets that can trigger chondrule-forming impacts 
is $\sim 0.1-1M_{\oplus}$ at $a\simeq 1-3$ AU.
This can result in $\Delta m_{ch}$ that becomes about $6 \times 10^{24}-10^{25}$ g.
Thus, even if planetary migration is taken into account,
it may still be marginally possible to obtain the value of $\Delta m_{ch}$ that is larger than the mass of the present asteroid belt.
It is however obvious that more detailed studies are needed to fully examine the effect of planetary migration.

For another dynamical process,
we discuss radial drift that is effective for mm-sized particles.
As found in the pioneering work of \citet{w77},
solid particles moving in gas disks experience head winds that arise from the disk gas.
The fundamental origin of such winds is the difference between the gas and the dust motion;
while the gas motion is slightly sub-keplerian due to its pressure,
dust is in purely keplerian motion.
It is well known that, under the action of head winds,
mm-sized particles lose their angular momentum so efficiently 
that they will be accreted onto the host stars very rapidly.
In order for chondrules formed via impact jetting in gas disks to be found in chondrites,
therefore, two possibilities can be considered.
The first possibility may be efficient accretion of chondrules onto planetesimals.
For this case,  it is required that most chondrules should be incorporated into forming planetesimals 
before chondrules experience significant radial drift.
While planetesimal formation itself is currently under active investigation \citep[e.g.,][]{cy10,jbt14},
one of the promising ideas, streaming instabilities, may be able to achieve this situation \citep[e.g.,][]{yg05,jomk07}.
The second possibility may be trapping of chondrules in turbulent eddies,
which can suppress efficient radial drift of chondrules \citep[e.g.,][]{chp01}.
Since it is far beyond the scope of this paper to discuss these two possibilities in detail,
we will leave a more dedicated investigation for future work.

One of the most plausible processes to change the timescale of forming chondrules is involved with disk turbulence.
In the above calculations, we have assumed laminar gas disks.
Nonetheless, it is well known that protoplanetary disks are quite turbulent \citep[e.g.,][]{wc11}.
When planetesimals move through turbulent disks, 
they experience random torque that originates from density fluctuations in the disks \citep[e.g.,][]{lsa04,jgm06}.
One may consider that the torque would be significant enough to pump up the eccentricity of planetesimals,
which can speed up the onset of chondrule forming impacts.
\citet{igm08} have examined this effect, and 
shown that random torque can lead to even collisional fragmentation of planetesimals 
due to a high value of their eccentricity.
This effect is prominent especially for smaller planetesimals that have $< 10-100$ km in size.
For larger planetesimals ($> 100$ km),
the damping effect that is caused either by gas disks or by inelastic collisions 
can suppress the effect of random torque. 
Thus, our results derived from laminar disks may be applicable even for the turbulent disk case.
Since the above argument depends on disk parameters such as the strength of disk turbulence,
we will perform a more quantitative analysis in our future work.

Another process that can be considered to accelerate chondrule formation may be the formation of nearby planets. 
The above results are obtained, 
assuming that the chondrule-forming site is essentially isolated from surrounding planet-forming regions.
This is highly idealized obviously, 
because multiple planets probably formed simultaneously in the solar nebula.
It can be anticipated that, when the formation of a massive planet such as Jupiter is ongoing,
the eccentricity of planetesimals in the chondrule-forming region can be pumped up,
which can accelerate chondrule formation.
In a subsequent paper, we will examine this effect by performing $N$-body simulations (Oshino et al. in prep).

\section{Conclusions} \label{conc}

Understanding of how chondrules formed can provide profound insights as to origins of our Solar system.
We have investigated a scenario of chondrule formation 
in which planetesimal collisions that occur during the formation of terrestrial planets and cores of gas giants, 
play a dominant role.
This scenario has recently been proposed by \citet{jmm15} which suggest that
planetesimal collisions and the resultant impact jetting 
may be efficient enough to reproduce the primordial abundance of chondrules.
They show that impact jetting leads to melting ejected materials 
when the impact velocity of planetesimals ($v_{imp}$) exceeds about 2.5 km s$^{-1}(\equiv v_c)$.

We have examined the scenario in detail.
To clarify how effective the scenario is to form chondrules,
we have adopted a simple semi-analytical approach that has been developed 
based on the results of more detailed $N$-body simulations \citep[e.g.,][]{im93,ki00,ki02}.
For pedagogical purpose, 
we have initially demonstrated how protoplanets form in a swarm of planetesimals (see Figure \ref{fig1}).
As shown by many previous $N$-body simulations,
the formation of protoplanets is regulated by the so-called runaway and oligarchic growth,
which is often called as planetary accretion.
The growth rate of protoplanets is then determined by the random velocity of field planetesimals that consist of a planetesimal disk
out of which protoplanets are born.

Armed with the theory of planetary accretion,
we have investigated two kinds of collisions and the resultant chondrule formation; 
protoplanet-planetesimal collisions and planetesimal-planetesimal ones.
For protoplanet-planetesimal collisions, 
the impact velocity of colliding planetesimals is determined mainly by the surface escape velocity (see equation (\ref{eq:v_imp})),
and as a growing protoplanet approaches the isolation mass ($M_{p,iso}$),
the value of $v_{imp}$ can reach $v_c$ (see Figure \ref{fig2}).
We have computed a fractional number of planetesimals ($F_{pl}(v_{imp}> v_c)$) that can satisfy the condition that $v_{imp} > v_c$,
and shown that the value of $F_{pl}(v_{imp}> v_c)$ becomes unity when $v_{imp} > v_c$.
The results therefore suggest that chondrule formation can occur efficiently at that time.
We have confirmed the trend by computing the cumulative mass of chondrules ($\Delta m_{ch,M-m}$) 
that can be formed by impact jetting via protoplanet-planetesimal collisions.
We have found that the value of $\Delta m_{ch,M-m}$ increases rapidly with time,
once $v_{imp}> v_c$.
It is important that $\Delta m_{ch,M-m}$ is characterized well by $F_{ch}M_{p,iso}$,
where $F_{ch}$ is a fractional mass of planetesimals that can eventually be converted to chondrules (see Table \ref{table1}).
For planetesimal-planetesimal collisions, 
we have shown that $v_{imp} < v_c$ for all the stages of planetary accretion (see Figure \ref{fig3}).
This arises from $v_{imp}$ that is regulated by the random velocity ($v_m$) of small planetesimals,
which can efficiently be damped by gas drag.
The largest contribution to chondrule formation occurs once protoplanets obtain $M_{p,iso}$ and the formation of protoplanets is finished.
This is because protoplanets can pump up the value of $v_m$ most efficiently at that time,
so that the value of $F_{pl}(v_{imp}>v_c)$ becomes highest.
Nonetheless, the resultant value of chondrule mass ($\Delta m_{ch,m-m}$) formed by planetesimal-planetesimal collisions is very small, 
compared with that of $\Delta m_{ch,M-m}$.
Thus, we have demonstrated that $\Delta m_{ch,M-m}$ can provide a major contribution to the abundance of chondrules and
the value can readily estimated if planetary accretion is completed within a given disk lifetime.

We have also examined the formation timing of chondrules.
This has been done by focusing both on when $v_{imp}$ becomes comparable to $v_c$ (i.e., $\tau = \tau_{int}$)
and on when protoplanets obtain $M_{p,iso}$ (i.e., $\tau=\tau_{end}$).
We have found that planetary accretion can complete within an upper limit of disk lifetimes ($\tau_d =10^7$ years)
at $a \la2$ AU for the standard disk ($f_d=1$)
and at $a \la 3$ AU for the massive disk ($f_d=3$) (see Figure \ref{fig4}).
These results have enabled further simplification of the estimation of $\Delta m_{ch,M-m}$ and $\Delta m_{ch,m-m}$,
wherein time-independent calculations have been adopted.
We have then investigated how both $\Delta m_{ch,M-m}$ and $\Delta m_{ch,m-m}$ behave 
as a function of the orbital distance ($a$) from the central star.
We have shown that, as long as oligarchic growth can be done within a given disk lifetime,
the time-independent calculations can reproduce the results of the time-dependent approach; 
$\Delta m_{ch,M-m} \simeq F_{ch}M_{p,iso}$ and $\Delta m_{ch,M-m} \gg \Delta m_{ch,m-m}$ (see Figure \ref{fig5}).
In addition, we have explored the effect of planetesimal mass ($m_{pl}$) on the total chondrule mass,
and shown that our results are valid for a wide range of $m_{pl}$ (see Figures \ref{fig6}, \ref{fig7}, and \ref{fig8}).
Thus, we can conclude that, 
based on a semi-analytical  formulation developed from the results of more detailed, direct $N-$body simulations,
planetary accretion can act as an efficient process to form chondrules,
which is consistent with the conclusion of \citet{jmm15}.

We have discussed an implication of impact jetting for chondrule formation.
Based on the above results, impact jetting is promising in the sense that 
the resultant abundance of chondrules is higher than the mass of the present asteroid belt (see Figure \ref{fig5}).
Nonetheless, more intensive work would be needed to examine 
whether or not the value is sufficient enough to reproduce chondrules that are currently found in chondrites.
As demonstrated above, when a value of disk mass and lifetime is given,
there is a clear upper limit on the the total mass of chondrules formed by impact jetting,
which is formulated by $F_{ch}M_{p,iso}$.
This estimate, however, is valid only if some dynamical processes are negligible 
such as planetary migration and radial drift of mm-sized particles.
In addition, the timing of chondrule formation can vary when other physical processes are included in the model.
These include disk turbulence and the formation of surrounding massive planets,
both of which can pump up the eccentricity of planetesimals and hence can speed up chondrule formation.

In a series of subsequent papers, we will investigate these effects and 
further examine how important impact jetting triggered by planetesimal collisions is to form chondrules.


\acknowledgments

The authors thank a referee, H. J. Melosh, for useful comments which substantially improve the quality of this work.
Numerical computations were in part carried out on GRAPE system at Center for Computational Astrophysics, 
National Astronomical Observatory of Japan.
Y.H. is supported by EACOA Fellowship that is supported by East Asia Core Observatories Association which consists of 
the Academia Sinica Institute of Astronomy and Astrophysics, the National Astronomical Observatory of Japan, the National Astronomical 
Observatory of China, and the Korea Astronomy and Space Science Institute.






\bibliographystyle{apj}          

\bibliography{apj-jour,adsbibliography}    

\begin{thebibliography}{54}
\expandafter\ifx\csname natexlab\endcsname\relax\def\natexlab#1{#1}\fi

\bibitem[{Adachi {et~al.}(1976)Adachi, Hayashi, \& Nakazawa}]{ahn76}
Adachi, I., Hayashi, C., \& Nakazawa, K. 1976, Prog. Theor. Phys., 56, 1756

\bibitem[{{Amelin} {et~al.}(2010){Amelin}, {Kaltenbach}, {Iizuka}, {Stirling},
  {Ireland}, {Petaev}, \& {Jacobsen}}]{akl10}
{Amelin}, Y., {Kaltenbach}, A., {Iizuka}, T., {Stirling}, C.~H., {Ireland},
  T.~R., {Petaev}, M., \& {Jacobsen}, S.~B. 2010, Earth and Planetary Science
  Letters, 300, 343

\bibitem[{{Baruteau} {et~al.}(2014){Baruteau}, {Crida}, {Paardekooper},
  {Masset}, {Guilet}, {Bitsch}, {Nelson}, {Kley}, \& {Papaloizou}}]{bcp14}
{Baruteau}, C., {Crida}, A., {Paardekooper}, S.-J., {Masset}, F., {Guilet}, J.,
  {Bitsch}, B., {Nelson}, R., {Kley}, W., \& {Papaloizou}, J. 2014, Protostars
  and Planets VI, 667

\bibitem[{Baruteau \& Masset(2008)}]{bm08}
Baruteau, C. \& Masset, F.~S. 2008, \apj, 672, 1054

\bibitem[{{Bitsch} {et~al.}(2014){Bitsch}, {Morbidelli}, {Lega}, \&
  {Crida}}]{bml14}
{Bitsch}, B., {Morbidelli}, A., {Lega}, E., \& {Crida}, A. 2014, \aap, 564,
  A135

\bibitem[{{Bottke} {et~al.}(2005){Bottke}, {Durda}, {Nesvorn{\'y}}, {Jedicke},
  {Morbidelli}, {Vokrouhlick{\'y}}, \& {Levison}}]{bdn05}
{Bottke}, W.~F., {Durda}, D.~D., {Nesvorn{\'y}}, D., {Jedicke}, R.,
  {Morbidelli}, A., {Vokrouhlick{\'y}}, D., \& {Levison}, H.~F. 2005, Icarus,
  179, 63

\bibitem[{Chiang \& Youdin(2010)}]{cy10}
Chiang, E. \& Youdin, A.~N. 2010, AREPS, 38, 493

\bibitem[{{Ciesla} \& {Cuzzi}(2006)}]{cc06i}
{Ciesla}, F.~J. \& {Cuzzi}, J.~N. 2006, Icarus, 181, 178

\bibitem[{{Connelly} {et~al.}(2012){Connelly}, {Bizzarro}, {Krot}, {Nordlund},
  {Wielandt}, \& {Ivanova}}]{cbk12}
{Connelly}, J.~N., {Bizzarro}, M., {Krot}, A.~N., {Nordlund}, {\AA}.,
  {Wielandt}, D., \& {Ivanova}, M.~A. 2012, Science, 338, 651

\bibitem[{{Cuzzi} {et~al.}(2001){Cuzzi}, {Hogan}, {Paque}, \&
  {Dobrovolskis}}]{chp01}
{Cuzzi}, J.~N., {Hogan}, R.~C., {Paque}, J.~M., \& {Dobrovolskis}, A.~R. 2001,
  \apj, 546, 496

\bibitem[{{Davis} {et~al.}(2014){Davis}, {Alexander}, {Ciesla}, {Gounelle},
  {Krot}, {Petaev}, \& {Stephan}}]{dac14}
{Davis}, A.~M., {Alexander}, C.~M.~O.~., {Ciesla}, F.~J., {Gounelle}, M.,
  {Krot}, A.~N., {Petaev}, M.~I., \& {Stephan}, T. 2014, Protostars and Planets
  VI, 809

\bibitem[{{DeMeo} \& {Carry}(2014)}]{dc14}
{DeMeo}, F.~E. \& {Carry}, B. 2014, \nat, 505, 629

\bibitem[{{Desch} \& {Cuzzi}(2000)}]{dc00}
{Desch}, S.~J. \& {Cuzzi}, J.~N. 2000, Icarus, 143, 87

\bibitem[{{Desch} {et~al.}(2012){Desch}, {Morris}, {Connolly}, \&
  {Boss}}]{dmc12}
{Desch}, S.~J., {Morris}, M.~A., {Connolly}, H.~C., \& {Boss}, A.~P. 2012,
  Meteoritics and Planetary Science, 47, 1139

\bibitem[{Goldreich \& Tremaine(1980)}]{gt80}
Goldreich, P. \& Tremaine, S. 1980, \apj, 241, 425

\bibitem[{{Hasegawa} \& {Pudritz}(2010)}]{hp10}
{Hasegawa}, Y. \& {Pudritz}, R.~E. 2010, \apjl, 710, L167

\bibitem[{{Hasegawa} \& {Pudritz}(2012)}]{hp12}
---. 2012, \apj, 760, 117

\bibitem[{Hayashi(1981)}]{h81}
Hayashi, C. 1981, Prog. Theor. Phys. Suppl., 70, 35

\bibitem[{{Helled} {et~al.}(2014){Helled}, {Bodenheimer}, {Podolak}, {Boley},
  {Meru}, {Nayakshin}, {Fortney}, {Mayer}, {Alibert}, \& {Boss}}]{hbp14}
{Helled}, R., {Bodenheimer}, P., {Podolak}, M., {Boley}, A., {Meru}, F.,
  {Nayakshin}, S., {Fortney}, J.~J., {Mayer}, L., {Alibert}, Y., \& {Boss},
  A.~P. 2014, Protostars and Planets VI, 643

\bibitem[{{Hewins} {et~al.}(2005){Hewins}, {Connolly}, \& {Libourel}}]{hcl05}
{Hewins}, R.~H., {Connolly}, Lofgren, G.~E. J. H.~C., \& {Libourel}, G. 2005,
  in Astronomical Society of the Pacific Conference Series, Vol. 341,
  Chondrites and the Protoplanetary Disk, ed. A.~N. {Krot}, E.~R.~D. {Scott},
  \& B.~{Reipurth}, 286

\bibitem[{{Hood} {et~al.}(2009){Hood}, {Ciesla}, {Artemieva}, {Marzari}, \&
  {Weidenschilling}}]{hca09}
{Hood}, L.~L., {Ciesla}, F.~J., {Artemieva}, N.~A., {Marzari}, F., \&
  {Weidenschilling}, S.~J. 2009, Meteoritics and Planetary Science, 44, 327

\bibitem[{Ida {et~al.}(2008)Ida, Guillot, \& Morbidelli}]{igm08}
Ida, S., Guillot, T., \& Morbidelli, A. 2008, \apj, 686, 1292

\bibitem[{Ida \& Makino(1993)}]{im93}
Ida, S. \& Makino, J. 1993, Icarus, 106, 210

\bibitem[{{Iida} {et~al.}(2001){Iida}, {Nakamoto}, {Susa}, \&
  {Nakagawa}}]{ins01}
{Iida}, A., {Nakamoto}, T., {Susa}, H., \& {Nakagawa}, Y. 2001, Icarus, 153,
  430

\bibitem[{{Johansen} {et~al.}(2014){Johansen}, {Blum}, {Tanaka}, {Ormel},
  {Bizzarro}, \& {Rickman}}]{jbt14}
{Johansen}, A., {Blum}, J., {Tanaka}, H., {Ormel}, C., {Bizzarro}, M., \&
  {Rickman}, H. 2014, Protostars and Planets VI, 547

\bibitem[{Johansen {et~al.}(2007)Johansen, Oishi, {Mac Low}, Klahr, Henning, \&
  Youdin}]{jomk07}
Johansen, A., Oishi, J.~S., {Mac Low}, M.-M., Klahr, H., Henning, T., \&
  Youdin, A. 2007, \nat, 448, 1022

\bibitem[{{Johnson} {et~al.}(2015){Johnson}, {Minton}, {Melosh}, \&
  {Zuber}}]{jmm15}
{Johnson}, B.~C., {Minton}, D.~A., {Melosh}, H.~J., \& {Zuber}, M.~T. 2015,
  \nat, 517, 339

\bibitem[{Johnson {et~al.}(2006)Johnson, Goodman, \& Menou}]{jgm06}
Johnson, E.~T., Goodman, J., \& Menou, K. 2006, \apj, 647, 1413

\bibitem[{{Jones} \& {Lofgren}(1993)}]{jl93}
{Jones}, R.~H. \& {Lofgren}, G.~E. 1993, Meteoritics, 28, 213

\bibitem[{Kley \& Nelson(2012)}]{kn12}
Kley, W. \& Nelson, R.~P. 2012, \araa, 50, 211

\bibitem[{Kokubo \& Ida(1996)}]{ki96}
Kokubo, E. \& Ida, S. 1996, Icarus, 123, 180

\bibitem[{Kokubo \& Ida(1998)}]{ki98}
---. 1998, Icarus, 131, 171

\bibitem[{Kokubo \& Ida(2000)}]{ki00}
---. 2000, Icarus, 143, 15

\bibitem[{Kokubo \& Ida(2002)}]{ki02}
---. 2002, \apj, 581, 666

\bibitem[{Kretke \& Lin(2012)}]{kl12}
Kretke, K.~A. \& Lin, D. N.~C. 2012, \apj, 755, 74

\bibitem[{Laughlin {et~al.}(2004)Laughlin, Steinacker, \& Adams}]{lsa04}
Laughlin, G., Steinacker, A., \& Adams, F.~C. 2004, \apj, 608, 489

\bibitem[{{Lissauer} \& {Stewart}(1993)}]{ls93}
{Lissauer}, J.~J. \& {Stewart}, G.~R. 1993, in Protostars and Planets III, ed.
  E.~H. {Levy} \& J.~I. {Lunine}, 1061--1088

\bibitem[{{Minton} \& {Levison}(2014)}]{ml14}
{Minton}, D.~A. \& {Levison}, H.~F. 2014, Icarus, 232, 118

\bibitem[{{Morfill} {et~al.}(1993){Morfill}, {Spruit}, \& {Levy}}]{msl93}
{Morfill}, G., {Spruit}, H., \& {Levy}, E.~H. 1993, in Protostars and Planets
  III, ed. E.~H. {Levy} \& J.~I. {Lunine}, 939--978

\bibitem[{{Morris} \& {Desch}(2009)}]{md09}
{Morris}, M.~A. \& {Desch}, S.~J. 2009, Astrobiology, 9, 965

\bibitem[{{Morris} \& {Desch}(2010)}]{md10}
---. 2010, \apj, 722, 1474

\bibitem[{{Ohtsuki} {et~al.}(1993){Ohtsuki}, {Ida}, {Nakagawa}, \&
  {Nakazawa}}]{oin93}
{Ohtsuki}, K., {Ida}, S., {Nakagawa}, Y., \& {Nakazawa}, K. 1993, in Protostars
  and Planets III, ed. E.~H. {Levy} \& J.~I. {Lunine}, 1089--1107

\bibitem[{Paardekooper \& Mellema(2006)}]{pm06}
Paardekooper, S.-J. \& Mellema, G. 2006, \aap, 459, L17

\bibitem[{{Raymond} {et~al.}(2014){Raymond}, {Kokubo}, {Morbidelli},
  {Morishima}, \& {Walsh}}]{rkm14}
{Raymond}, S.~N., {Kokubo}, E., {Morbidelli}, A., {Morishima}, R., \& {Walsh},
  K.~J. 2014, Protostars and Planets VI, 595

\bibitem[{{Sanders} \& {Scott}(2012)}]{ss12}
{Sanders}, I.~S. \& {Scott}, E.~R.~D. 2012, Meteoritics and Planetary Science,
  47, 2170

\bibitem[{{Shu} {et~al.}(2001){Shu}, {Shang}, {Gounelle}, {Glassgold}, \&
  {Lee}}]{ssg01}
{Shu}, F.~H., {Shang}, H., {Gounelle}, M., {Glassgold}, A.~E., \& {Lee}, T.
  2001, \apj, 548, 1029

\bibitem[{{Shu} {et~al.}(1996){Shu}, {Shang}, \& {Lee}}]{ssl96}
{Shu}, F.~H., {Shang}, H., \& {Lee}, T. 1996, Science, 271, 1545

\bibitem[{Thommes {et~al.}(2003)Thommes, Duncan, \& Levison}]{tdl03}
Thommes, E.~W., Duncan, M.~J., \& Levison, H.~F. 2003, Icarus, 161, 431

\bibitem[{{Urey} \& {Craig}(1953)}]{uc53}
{Urey}, H.~C. \& {Craig}, H. 1953, \gca, 4, 36

\bibitem[{Weidenschilling(1977)}]{w77}
Weidenschilling, S.~J. 1977, \mnras, 180, 57

\bibitem[{Wetherill \& Stewart(1989)}]{ws89}
Wetherill, G.~W. \& Stewart, G.~R. 1989, Icarus, 77, 330

\bibitem[{Williams \& Cieza(2011)}]{wc11}
Williams, J.~P. \& Cieza, L.~A. 2011, \araa, 49, 67

\bibitem[{{Wood}(1963)}]{w96}
{Wood}, J.~A. 1963, Icarus, 2, 152

\bibitem[{Youdin \& Goodman(2005)}]{yg05}
Youdin, A.~N. \& Goodman, J. 2005, \apj, 620, 459

\end{thebibliography}

\end{document}